\documentclass[secnumarabic, graphics,floatfix, nofootinbib,tightenlines,nobibnotes, aps, 12pt]{revtex4}
\usepackage{graphicx}
\usepackage[english]{babel}
\usepackage[utf8]{inputenc}
\usepackage{amsmath,amssymb}
\usepackage{amsfonts}
\usepackage{multirow}
\usepackage{booktabs}
\usepackage[section]{placeins}
\usepackage{float}%
\begin{document}

%%%%%%%%%%%%%%%%%%%%%%%%%%%%%%%%%%%%%%%%%%%%%%%%%%%%%%%%%%%%%%%
\newcommand{\bq}{\begin{equation}}
\newcommand{\eq}{\end{equation}}
\newcommand{\bqn}{\begin{eqnarray}}
\newcommand{\eqn}{\end{eqnarray}}
\newcommand{\nb}{\nonumber}
\newcommand{\lb}{\label}
%
%\newcommand{\PRL}{Phys. Rev. Lett.}
%\newcommand{\PL}{Phys. Lett.}
%\newcommand{\PR}{Phys. Rev.}
%\newcommand{\CQG}{Class. Quantum Grav.}
%%%%%%%%%%%%%%%%%%%%%%%%%%%%%%%%%%%%%%%%%%%%%%%%%%%%%%%%%%%%%%%

\title{The initial structure of chondrule dust rims I: electrically neutral grains }

\author{C. Xiang$^{1}$}%\email{}
\author{A. Carballido$^{1}$}
\author{R.D. Hanna$^{2}$}
\author{L.S. Matthews$^{1}$}
\author{T.W. Hyde$^{1}$}
\affiliation{$^{1}$ Center for Astrophysics, Space Physics, and Engineering Research, Baylor University, Waco, TX 76798-7316, USA;}
\affiliation{$^{2}$ Jackson School of Geosciences, University of Texas, Austin, TX 78712, USA}

\begin{abstract}

In order to characterize the early growth of fine-grained dust rims (FGRs) that commonly surround chondrules, we perform numerical simulations of dust accretion onto chondrule surfaces. We employ a Monte Carlo algorithm to simulate the collision of dust monomers having radii between 0.5 and 10 $\mu$m with chondrules whose radii are between 500 and 1000 $\mu$m, in 100-$\mu$m increments. The collisions are driven by Brownian motion and solar nebula turbulence. After each collision, the colliding particles either stick at the point of contact, roll or bounce. We limit accretion of dust monomers (and in some cases, dust aggregates) to a small patch of the chondrule surface, for computational expediency. We model the morphology of the dust rim and the trajectory of the dust particle, which are not considered in most of the previous works. Radial profiles of FGR porosity show that rims formed in weak turbulence are more porous (with a porosity of 60-74\%) than rims formed in stronger turbulence (with a porosity of 52-60\%). The lower end of each range corresponds to large chondrules and the upper end to small chondrules, meaning that the chondrule size also has an impact on FGR porosity. The thickness of FGRs depends linearly on chondrule radius, and the slope of this linear dependency increases with time, and decreases with the turbulence strength. The porosity of FGRs formed by dust aggregates is $\sim 20\%$ on average greater than that of FGRs formed by single monomers. In general, the relatively high porosities that we obtain are consistent with those calculated by previous authors from numerical simulations, as well as with initial FGR porosities inferred from laboratory measurements of rimmed chondrule samples and rimmed chondrule analogs. 
%In this work, we use a molecular dynamics code to model the growth of fine-grained chondrule rims through the collection of micron-sized dust grains. We examine the time elapsed to collect rims of a certain thickness and the rim porosity as a function of the conditions present in a protoplanetary disk, such as turbulence strength, and the chondrule size. In addition, we compare rims formed by the collection of single spherical grains and rims formed through the collection of aggregate dust grains. As expected, chondrule rims formed in strongly turbulent environments are more compact and grow more rapidly than those formed in weak turbulence. The chondrule fabric records a range of information during its formation in the nebula, and the investigation of the initial rim material, before it is processed, may give an indication of the environmental conditions in a protoplanetary disk.

\end{abstract}

%\textbf{Keywords}: Black hole, Quasinormal modes, Matrix Method, Taylor Series, Eigenvalue
\maketitle
%%%%%%%%%%%%%%%%%%%%%%%%%%%%%%%%%%%%%%%%%%%%%%%%%%%%%%%%%%%%%
\section{Introduction}
%%%%%%%%%%%%%%%%%%%%%%%%%%%%%%%%%%%%%%%%%%%%%%%%%%%%%%%%%%%%%
\renewcommand{\theequation}{1.\arabic{equation}} \setcounter{equation}{0}

%In a gas- and dust- rich protosolar nebula, the motion of dust grains is driven by a variety of processes such as Brownian motion, turbulence, radial drift and settling towards the mid-plane. The relative motion between dust grains leads to collisions, and the dust gradually forms fluffy dust balls (Cuzzi et al. 2006). High temperature events, such as direct radiation from the protosun, the occurrence of shock fronts resulting from gravitational instabilities (Boss et at, 2005) and collisions between planetesimals, can rapidly heat the dust aggregates, melting them into droplets that quickly cool into sub-millimeter-sized quasi-spherical chondrules composed of glass and minerals (Cuzzi et al. 2006). By studying the textures, mineral compositions and isotopic effects of chondrule samples, we can interpret their thermal histories and gain information about the formation of the solar system.

Chondrules, the formerly molten, quasi-spherical, (sub)millimeter-sized silicate grains that are the primary constituents of chondritic meteorites, are commonly surrounded by fine-grained dust rims (FGRs). Along with refractory components such as calcium-aluminum inclusions, and a matrix made up of both crystalline and amorphous grains, rimmed chondrules encase valuable information regarding early processes in our solar system. In particular, chondrule FGRs may have acted as a ``glue" that facilitated accretion of many rimmed chondrules into centimeter-sized objects, possible precursors to asteroids (Ormel et al. 2008). 

 FGRs have been observed in both optical and scanning electron microscopy studies of carbonaceous chondrites, and are particularly visible in CM chondrites (Ashworth 1977, Metzler et al. 1992, Brearley 1993). The origin of FGRs has been somewhat disputed, with some researchers proposing that FGRs formed in the parent body environment, either by attachment and compaction of dust onto chondrules in regolith (Sears et al. 1993, Trigo-Rodriguez et al. 2006, Takayama and Tomeoka 2012), or through aqueous alteration of chondrules (Sears et al. 1993, Takayama and Tomeoka 2012). However, several lines of evidence suggest that FGRs formed in a nebular setting, before the rimmed chondrules were incorporated into their parent bodies (Metzler et al. 1992, Morfill et al. 1998, Brearley 1999). The presence of pre-solar grains in CR chondrite FGRs led Leitner et al. (2016) to conclude that those rims had a nebular origin, since pre-solar silicate and oxide abundances in the rims differ from those in the interchondrule matrix, indicating different alteration paths of both meteoritic components. Bland et al. (2011) mapped the orientation of submicron grains in the Allende CV chondrite, and calculated an initial rim porosity of 70-80\% by relating fabric intensity to net compression. Such high porosity values are similar to those obtained from Monte Carlo simulations (Ormel et al. 2008) and laboratory experiments (Beitz et al. 2013), which assume nebular conditions. More recently, using X-ray computed tomography, Hanna and Ketcham (2018) examined the 3D morphology of FGRs in the CM chondrite Murchison, and found a power law relation between FGR volume and chondrule radius, consistent with rim accretion in a weakly turbulent solar nebula as calculated by Cuzzi (2004). 

The significance of a nebular scenario for FGR formation can not be overstated. A possible path towards the emergence of asteroidal parent bodies, composed of agglomerates of rimmed chondrules, could involve an essentially hydrodynamic process: the runaway convergence of chondrules due to the relative drift between the solar nebula gas and small solids (Carrera et al. 2015). This streaming instability, as it is known, has its origin in the radial pressure gradient that supports the nebular gas, but not the solid component. If chondrules acquired dust envelopes while being suspended in the solar nebula, the resulting rimmed chondrules could have formed dense clumps due to the streaming instability. These clumps, in turn, would have facilitated low-velocity sticking between rimmed chondrules.

Previous modeling approaches to the problem of FGR formation are few but varied. Morfill et al. (1998) integrated an evolution equation for the radius of a rimmed chondrule, assumed to be suspended in a dusty-gaseous medium, and using a prescribed sticking efficiency between dust and chondrule. Cuzzi (2004) used a semi-analytical model to calculate rim volume as a function of chondrule volume under constant and variable dust densities. Ormel et al. (2008) employed a sophisticated Monte Carlo method to study inter-chondrule sticking in parameterized turbulence. They modeled the compaction of porous dust layers based on experimental and theoretical results from the physics of dust collisions (Chokshi et al. 1993, Dominik and Tielens 1997, Blum and Schrapler 2004), but without resolving the actual rim structure. Carballido (2011) performed simulations of dust sweep-up by chondrules in a local, magnetohydrodynamic (MHD), turbulent model of the solar nebula (SN), following the approach of Morfill et al. (1998). However, since the turbulence in Carballido (2011) was generated by the magnetorotational instability under \textit{ideal} MHD conditions, the turbulent regime needs to be revised in light of recent results indicating that \textit{non}-ideal MHD effects lead to much weaker turbulence than in the ideal MHD case (Bai \& Stone 2013, Bai 2014).

Here we develop a molecular dynamics model to simulate collisions between dust and chondrules, taking into consideration detailed collisional physics. For computational expediency, we assume that the chondrule accretes dust isotropically, and we restrict our study to a small patch on the chondrule surface. We track the evolution of the dust accretion process, and examine the structure of the resulting dust rim (or more precisely, \textit{partial} rim). All theoretical models of chondrule rim growth have assumed that the dust grains comprising the rims are electrically neutral, and we follow that assumption here. In reality, dust grains become charged to varying degrees in the radiative plasma environment of the solar nebula (Okuzumi et al. 2009, Matthews et al. 2012). We will study the effects of electric charge on the formation of FGRs in a follow-up paper. 

%We introduce electric charge on the dust in some of our simulations. To our knowledge, this is the first numerical study of FGR formation that takes into account electrical charging of dust grains. The trajectories of colliding dust grains can be altered by the electrostatic force acting between the grains, affecting their coagulation probability as well as their impact velocity (Matthews et al. 2012). The electrostatic force would therefore influence the structure of the dust rim, as well as the time scale of rim formation. The effect of dust charge on this process would be modified, in turn, by the turbulence strength, as dust grains entrained in SN regions with relatively strong turbulence would have a greater probability of overcoming the electrostatic barrier, and hence reaching the rimmed chondrule surface. For these reasons, it is important to take into account grain charge in models of FGR formation. If the nebular hypothesis is correct, the resulting FGR structure could ultimately be used to infer values of gas velocities, turbulent viscosity, and ionization state of the solar nebula (Ormel et al. 2008; Matthews et al. 2012; Okuzumi et al. 2009). 

Section 2 presents an overview of the relative velocities and collision outcomes for grains embedded in turbulent gas flow. The numerical methods for modeling the collection of dust on the chondrule surface are described in Section 3. We investigate the effects of turbulence strength and chondrule size on rim growth, as well as the difference between rims formed by accretion of dust monomers (particle aggregation) and rims formed by accretion of dust aggregates (cluster aggregation). The results analyzing the structure, porosity and the time to build the dust rims are presented in Section 4. Discussion of the results takes place in Section 5, and the main conclusions are summarized in Section 6.

%%%%%%%%%%%%%%%%%%%%%%%%%%%%%%%%%%%%%%%%%%%%%%%%%%%%%%%%%%%%
\section{BRIEF OVERVIEW OF DUST TURBULENT KINEMATICS AND COLLISION OUTCOMES
}\label{sec:kin}
%%%%%%%%%%%%%%%%%%%%%%%%%%%%%%%%%%%%%%%%%%%%%%%%%%%%%%%%%%%%%
\renewcommand{\theequation}{2.\arabic{equation}} \setcounter{equation}{0}

The initial structure of FGRs is determined by interactions between micrometer-sized dust grains and mm-sized chondrules, and also between the dust grains themselves. Small particles entrained in a turbulent gas flow develop relative velocities due to the difference in their coupling times with the gas. The kinetic energy of a dust particle approaching a chondrule is thus influenced by the turbulence strength. The collision energy determines the collision outcome: incoming dust particles can stick at the point of contact, cause restructuring of the dust rim, or bounce. At the same time, the relative velocities between chondrules and dust particles of various sizes will also determine the collision rate, which in turn determines the time needed to build up a rim of a certain thickness. 

\begin{figure*}[!htb]
\includegraphics[width=9cm]{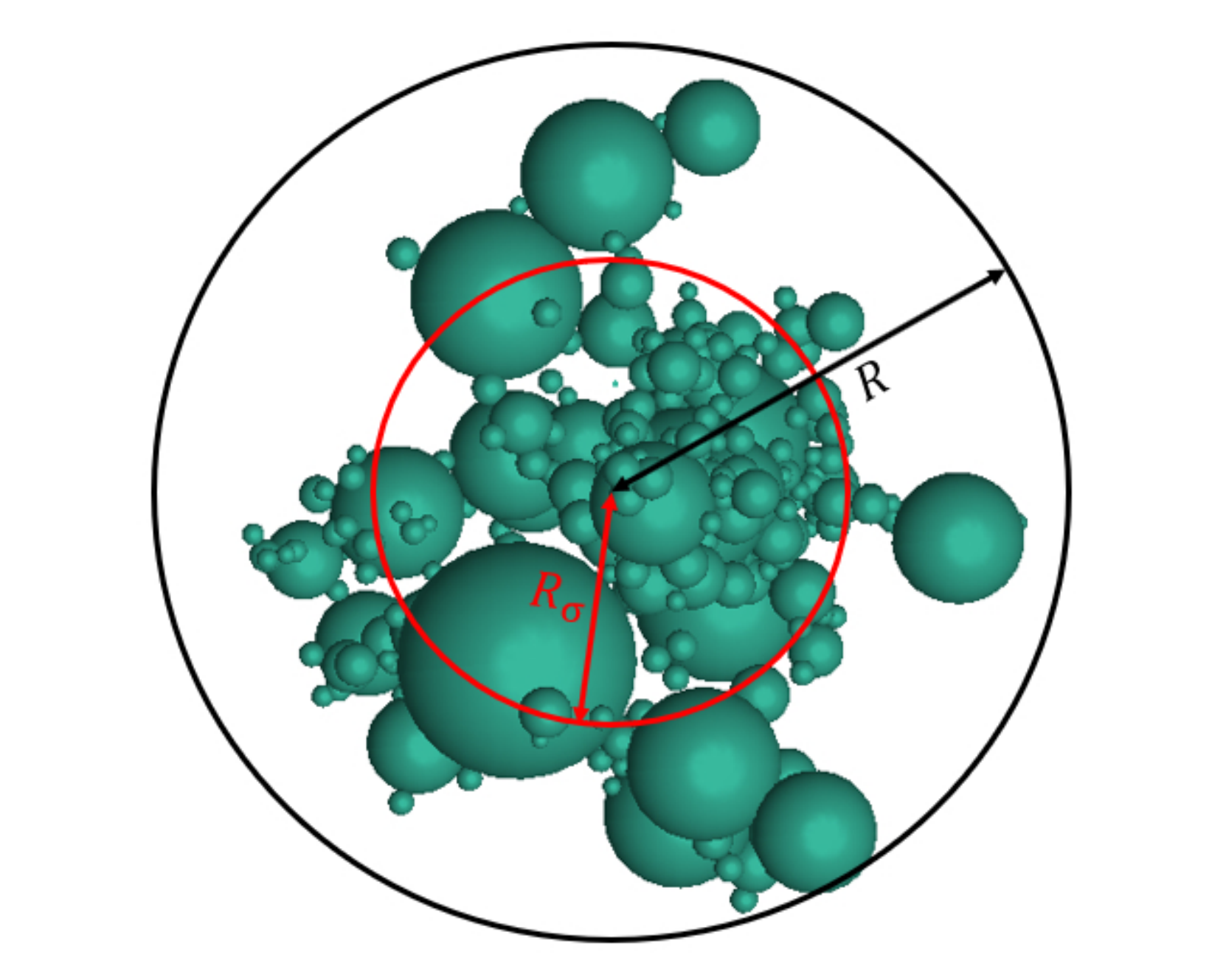}\includegraphics[width=7cm]{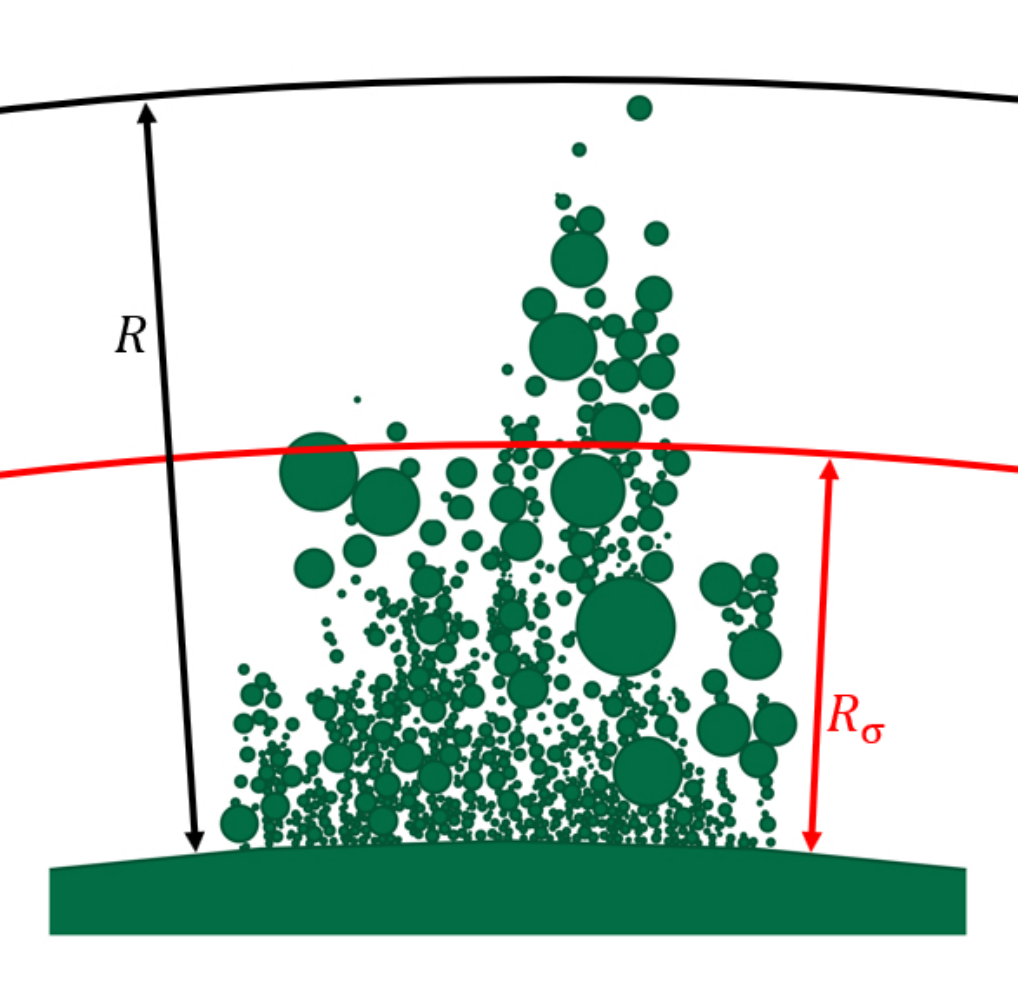}
\caption{Illustration of physical radius and equivalent radius for a) a dust aggregate, and b) a dust rim (shown is the monomer distribution on a vertical plane cutting through the center of a dust pile on the chondrule surface). The black arrow indicates the physical radius $R$, defined as the maximum radial extent from the center of mass for a) or the chondrule surface for b). The red arrow indicates the equivalent radius $R _{\sigma}$, as defined in the text. $R$ and $R _{\sigma}$ for a rimmed chondrule are equal to $R$ and $R _{\sigma}$ of the dust rim plus the radius of the chondrule core. }
\label{fig0}
\end{figure*}

In this work, two definitions of radius, equivalent radius $R _{\sigma}$ and physical radius $R$, are used to characterize the structure of dust aggregates and rimmed chondrules. The equivalent radius of a dust aggregate is the radius of a circle with area equal to the projected cross-section of the aggregate averaged over many orientations. The equivalent radius of a rimmed chondrule is defined as the sum of the radius of the chondrule core and the thickness of the inner region of the dust rim with a porosity lower than 70\%. For both the dust aggregate and the rimmed chondrule, the physical radius is the maximum radial extent from the center of mass (COM). The comparison of $R _{\sigma}$ and $R$ is indicated in figure \ref{fig0}.

In the rest of this section we briefly describe the type of dust motion and collision outcomes that we consider in our simulations. 

\subsection{Motion of dust in a protoplanetary disk}\label{sec:motion}

Various mechanisms impart relative velocities to solid particles in a protoplanetary disk (PPD), such as Brownian motion, inward radial drift, vertical settling towards the midplane, and turbulence (Brauer et al. 2008; Weidenschilling 1977; Voelk et al. 1980; Ormel and Cuzzi 2007; Ormel et al. 2008). The dominant source of relative velocities depends on the disk temperature and location, as well as on the particle properties, i.e., mass and porosity (Krijt et al. 2014). In this study, we only consider the turbulent and Brownian contributions, assuming the particles are well coupled to turbulent eddies, as systematic relative velocities due to drift and settling are much lower for particles sizes relevant to our problem (Rice et al. 2004; Dullenmond et al. 2004; Ormel et al. 2008). In this case, the total relative speed between a dust particle and chondrule is given by
\bqn
\lb{1}
v_{r}=\sqrt{v_{B}^{2}+v_{T}^{2}}
\eqn
where $ v_{B}$ is the relative velocity due to Brownian motion, 
\bqn
\lb{2}
v_{B}=\sqrt{\frac{8(m_{1}+m_{2})k_{B}T}{\pi m_{1}m_{2}}}
\eqn
with $ m_{1}$, $ m_{2}$ the masses of the two colliding particles, $k_{B}$ Boltzmann's constant, and with the gas temperature, $T=\frac{280}{\sqrt{r}}$, based on the minimum mass solar nebula (MMSN) model (Weidenschilling 1977; Hayashi 1981; Thommes et al. 2006), where $r$ is the heliocentric distance (which we take as 1 AU in this study).

A closed-form analytical expression for the relative turbulent velocity $ v_{T}$ was presented by Ormel and Cuzzi (2007) and Ormel et al. (2008), by comparing the stopping time $\tau_1=\frac{3}{4c_{g}\rho _{g}}\frac{m_1}{\pi a_1^{2}}$ of the largest particle (with $\rho_{g}$ the gas density, $c_{g}$ the gas thermal speed, and $m_1$ and $a_1$ the mass and equivalent radius of the particle), with the turn-over times of the largest and smallest turbulent eddies, $t_L$ and $t_s$:
\bqn\label{eq:vturb}
\lb{3}
v_{T}=\left\{\begin{matrix}
v_gRe^{1/4}( \rm{St}_{1}- St_{2})~&\text{for}&~\tau_1<t_s\\
v_g[2y_a-(1+\epsilon)+\frac{2}{1+\epsilon}(\frac{1}{1+y_a}+\frac{\epsilon ^{3}}{y_a+\epsilon })]^{1/2}\rm{St}_{1}^{1/2}~&\text{for}&~5t_s\simeq\tau_1\lesssim t_L\\
\left(\frac{1}{1+\rm{St}_1}+\frac{1}{1+\rm{St}_2}\right)~&\text{for}&~\tau_1\geq t_L
\end{matrix}\right.
\eqn
In this expression, $v_g$ is the gas speed; $Re$ is the Reynolds number, defined as the ratio of the turbulent viscosity, $\nu_{T}=\alpha c_{g}^{2}/\Omega$, to the molecular viscosity of gas, $\nu_{m}=c_{g}\lambda/2$ (Cuzzi et al. 1993), with $\alpha$ the turbulence strength (Shakura \& Sunyaev 1973), $\Omega$ the local Keplerian angular speed, and $\lambda$ the gas mean free path; the turn-over times of the largest and smallest eddies are $t_{L}=1/\Omega$ and $t_{s}=Re^{-1/2}t_{L}$, respectively; the Stokes numbers $\mathrm{St}_i$, $i=1,2$, are the ratios $\tau_i/t_L$; $y_{a}$ is a numerical constant taken to be 1.6; and the quantity $\epsilon$ is the ratio $ \frac{St_{1}}{St_{2}}\leq 1$ (Ormel et al. 2008).

For the regions of the solar nebula and particle sizes that we are considering, $v_T$ is determined by the first condition in Eq. (\ref{eq:vturb}), and the relative velocities depend on the size difference between the chondrule and the dust grain.

\subsection{Collision outcomes}\label{sec:outcomes}

For low-velocity collisions between $\mu$m- sized grains and (sub)mm-sized chondrules, i.e., $v<10$ cm s$^{-1}$, almost all collisions result in sticking at the point of contact (Ormel et al. 2008). However, collisions with energies that exceed a certain minimum threshold will result in restructuring, bouncing, fragmentation or mass transfer (Wurm et al. 2005; Kothe et al. 2010) [we note that more collisional outcomes, up to nine, are possible, as reported by G\"{u}ttler et al. (2010). For example, experiments show that bouncing can occur bimodally, either with compaction or with mass transfer. Nevertheless, for simplicity, we limit the outcomes to restructuring and bouncing]. Restructuring occurs when particles roll along the surface. For micron-sized particles, the main source of friction comes from bonds between atoms at the surface. New contacts can be made and old contacts can be broken only in step sizes of at least one atom (Dominik and Tielens 1997). The critical energy required to roll a distance equal to the size of an atom is 
\bqn\label{equaroll}
\lb{7}
e_{roll}=6\pi \gamma \xi_{\rm{crit}}^{2}
\eqn
where $\gamma$ is the surface energy per unit area and $ \xi_{\rm{crit}}$ is the average distance over which energy is dissipated. Following Dominik and Tielens (1997), we set $\xi_{\rm{crit}} = 1$ $\AA$. In order to cause non-negligible restructuring, the particle has to roll a finite distance. Therefore, we define the quantity $E_{roll}$, which is the energy required to roll a distance of 1000 $\AA$, 
\bqn
\lb{7}
E_{roll}=1000 e_{roll}
\eqn

This energy is independent of the size of the colliding particles, and rolling continues until all the energy is dissipated. The mechanism of energy dissipation that we employ during restructuring is described in Appendix A.

The critical bouncing velocity between two colliding grains is determined by the radii, surface energy, Poisson ratio, Young's modulus and density of the particles:
\bqn
\lb{7}
v_{cri}\simeq 3.86\frac{\gamma ^{5/6}}{E^{1/3}r^{5/6}\rho ^{1/2}}
\eqn
where $E$, $r$ and $\rho $ are the material constant (a function of the Poisson ratios and Young's moduli), the reduced radius of the two spheres, and the density of the grains, respectively. The detailed derivation is presented in Dominik and Tielens (1997).

%With the introduction of charged particles, the electrostatic interaction exerts a force and a torque on the incoming particle, causing it to decelerate (for like-charged particles) and rotate. The deceleration and deflection can cause the dust particle to miss the chondrule or reduce the amount of restructuring. Surface charging of dust grains is described in Section 3.3.

%%%%%%%%%%%%%%%%%%%%%%%%%%%%%%%%%%%%%%%%%%%%%%%%%%%%%%%%%%%%
\section{ NUMERICAL TREATMENT OF RIM GROWTH}
%%%%%%%%%%%%%%%%%%%%%%%%%%%%%%%%%%%%%%%%%%%%%%%%%%%%%%%%%%%%%
\renewcommand{\theequation}{3.\arabic{equation}} \setcounter{equation}{0}

As mentioned in the introduction, we only consider dust accretion onto a small area of a chondrule's surface, thus greatly expediting our calculations. In our treatment of chondrule rim growth, the factors that affect the coagulation process are the probability that two particles travel towards each other (determined by their cross-sectional area and relative velocity) and the type of interaction between them, which determines the collision outcome (i.e., sticking, bouncing, etc.). We use a combination of a Monte Carlo method and an N-body code to model these two factors.

Each simulation begins with a (sub)-millimeter-sized spherical chondrule as a target, with a radius between 500 $\mu$m and 1000 $\mu$m, placed at the origin [depending on chondrite group, chondrule diameters range from 100 $\mu$m to 2000$\mu$m, with approximate log-normal distributions (Friedrich et al. 2015)]. The potentially colliding dust grains are selected from a population of 10,000 dust particles with radii ranging from 0.5 to 10 $\mu$m, with a power law size distribution  $n(r)dr\propto r^{-3.5}dr$ (Mathis et al. 1977). In general, when two solid particles are far away from each other, their relative velocity depends largely on the particle sizes, as motion is driven by coupling of the solids to the gas. We use a Monte Carlo algorithm to randomly select dust particles that will collide with the chondrule, as well as determine the elapsed time interval between collisions, as described in Section \ref{sec:mc}. At close approach, the detailed collision process is modeled using an N-body algorithm, Aggregate Builder (AB), to determine the collision outcome, as well as any restructuring of the chondrule rim, as described in Section \ref{sec:AB}. %The interaction between charged particles at close approach is also taken into account using this algorithm.  The process for calculating the charge distribution on the dust aggregates and the chondrule is detailed in Section 3.3.  

At the beginning of the simulation, a population of dust particles with a range of sizes is grouped into 100 logarithmic bins by their radii, and the collision rates $ C_{ch,d}$, where $ch$ stands for chondrule and $d$ stands for dust particle, are initialized using the average radii and masses of particles in each bin. In each iteration, time is advanced by a random interval to the time when the next interaction between the chondrule and dust particles will occur. Then, a dust particle is randomly chosen to be shot towards the chondrule surface, based on the collision probabilities, with the collision outcome modeled by AB. 
\\

\subsection{Monte Carlo algorithm}\label{sec:mc}

The Monte Carlo algorithm is a mathematical method used to simulate the stochastic accretion of dust on the chondrule surface. The fundamental postulate of this algorithm is that there exists a function $ C_{ij}(i, j) d\tau $ which represents the probability that a given pair of particles $i$ and $j$ will coagulate in the time interval $ d\tau $. In our model, the collision is between the target chondrule $ch$ and a randomly chosen dust particle $d$. Particles with larger radii and larger relative velocity have a higher chance to collide. The volume that particle $d$ sweeps out relative to the chondrule $ch$ per unit time is $\sigma _{ch,d}\Delta v_{ch,d}$, where $\sigma _{ch,d}= \pi(r_{ch}+r_{d})^{2}$ is the collision cross section with $r_{ch}$ and $r_{d}$ the equivalent radii of the dust particle and the chondrule, and $\Delta v_{ch,d}$ is the relative velocity of the dust with respect to the chondrule. The ratio of this volume to the simulated volume $V$ in which the particles reside is proportional to the probability that the two particles collide per unit time,
\bqn
\lb{10}
C_{ch,d}=\sigma _{ch,d}\Delta v_{ch,d}/V.
\eqn

The first step in this method is to determine the random time interval between the collection of dust particles, consistent with the collision probabilities $ C_{ch,d}$ (e.g. see Gillespie 1975).  The probability per unit time of collecting any dust particle is 
\bqn
\lb{10}
C_{tot}=\sum_{j=1}^{N}C_{ch,d}
\eqn
with $N$ the total number of dust particles. During a given time interval $\tau$, the probability of collecting any dust particle is proportional to
\bqn
\lb{10}
P=exp[-C_{tot} \tau]
\eqn
The time elapsed between two collisions, according to this probability, is given by $ \tau=-ln(r_{1})/C_{tot}$ , with $r_{1}$ a random number between zero and one.  A second random number $r_{2}$ is used to select the incoming dust particle \textit{d} by finding the smallest integer satisfying (Gillespie 1975)
\bqn
\lb{10}
\sum_{i=1}^{d}C_{i,ch}> r_{2}C_{tot}
\eqn
In order to reduce the computational cost in calculating the collision probabilities $ C_{ch,d}$, the range of equivalent radii is divided into 100 logarithmic intervals, and particles of similar size (within the same interval) are binned into the same group. The average equivalent radius of each bin is used to calculate $\widetilde{C_{ch,d}}$, the collision rate between particles in the group $d$ and the chondrule $ch$:
\bqn\label{eq:totcij}
\lb{13}
\widetilde{C_{ch,d}}=g_{d}C_{ch,d}
\eqn
where $g_{d}$ is the number of particles in group $d$ (Ormel et at. 2007). After a dust particle from group $d$ collides with the chondrule, $g_{d}$ is decreased by 1. At the same time, a particle in the population is randomly chosen to be duplicated, and the number of particles in its group is increased by 1, so that the total number of particles stays the same during the simulation. The abstract volume $V$ is rescaled after each duplication procedure, in order to keep the dust spatial density constant.
\\

According to the power law size distribution that we employ to model dust particles, the initial population contains a small number of particles with large radii. In order to reduce the fluctuation caused by small number statistics, instead of creating monomers randomly based on the power law at the beginning of the simulation, we create monomers with evenly spaced radii within the radius range 0.5 $\mu$m $\leqslant r\leqslant$ 10 $\mu$m, and add weights to particles of different sizes according to the power law distribution. If particles in the $d^{th}$ group have weights $w_{1}, w_{2}, ... w_{k}$, then $g_{d}$ in (\ref{eq:totcij}) equals $\sum_{x=1}^{x=k}w_{x}$. The physical meaning of a non-integer weight is explained in Appendix B.
\\

\subsection{Aggregate Builder}\label{sec:AB}

Once the dust particle is selected, the detailed interaction is modeled using an $N$-body code, Aggregate Builder (Matthews et al. 2012), that takes into account the morphology of the dust rim and the trajectory of the incoming particle. 

\begin{figure*}[!htb]
\includegraphics[width=14cm]{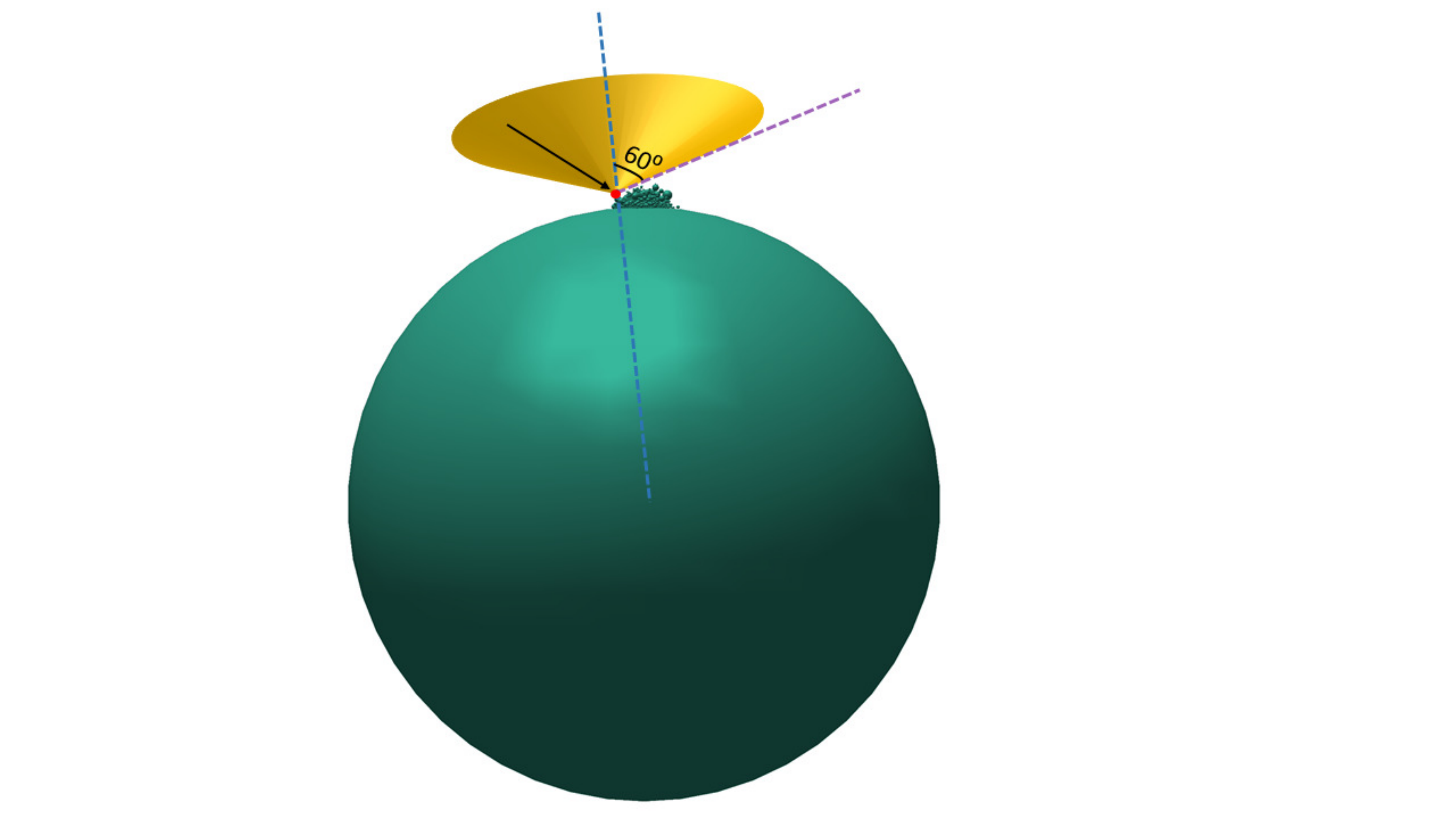}
\caption{Schematic showing the geometry of the collision between a dust particle and the chondrule surface. The green sphere represents the chondrule core with a dust pile on the surface, and the red spot is a randomly selected point on the dust pile. The yellow cone has its vertex at the selected point and its axis parallel to the normal of the surface of chondrule core at that point, with an angle of $60^{\circ}$ between the cone axis (blue dashed line) and the cone edge (purple dashed line). The incoming dust particle can be shot towards the point from any direction within the cone, with the black arrow as an example.}
\label{f0}
\end{figure*}

The chondrule is placed as a target with its center of mass at the origin. For computational expediency, we restrict dust particles to accumulate on a circular patch 100 $\mu$m in diameter on the chondrule surface. In each iteration, one dust particle is selected randomly from the dust population according to the previously computed collision rates, and is shot towards a randomly selected point on the target area from a random direction. The angle between this direction and the normal to the patch at the selected point is uniformly distributed between $0^{\circ}$ and $60^{\circ}$, so that it is less likely that dust particles impact the side of the dust pile, as illustrated in Fig. \ref{f0}. The initial distance between the chondrule and the incoming dust particle is set to be $2.5R_{\rm{ch}}$, with $R_{\rm{ch}}$ the radius of the chondrule. The initial relative velocity between the chondrule and the dust grains is set assuming that the dust is coupled to the gas in the solar nebula, as described in Sec. \ref{sec:motion}. %The trajectory and orientation of the incoming dust particle is then calculated  based on the electrostatic interactions, if present, to determine the collision outcome. 

The collision outcomes include sticking at the point of contact, rolling on the surface (which results in compaction), bouncing or a missed collision.
The energy required to initiate rolling ($e_{roll}$) is much smaller than the energy required for breaking a contact, while the energy $E_{roll}$ needed to initiate rolling is similar to that for breaking a contact. Therefore, it is easy for the particles to start rolling for a small distance, but difficult for them to roll a large distance without breaking the contact. The criteria for the different outcomes are (a) for sticking at the point of contact: the relative velocity is lower than the critical rolling velocity; (b) for rolling: the relative velocity is greater than the critical rolling velocity (determined by the critical rolling energy as in Eq.(\ref{equaroll}) and the mass of the incoming particle) and lower than the critical bouncing velocity; (c) for bouncing: the relative velocity is greater than the critical bouncing velocity; (d) for missing: particles are moving away from each other. In our simulation, particles have low relatively velocities and no bouncing is detected.

%As a dust particle approaches the chondrule surface or a rim particle, a collision check determines if the two collliding particles are separated by a distance smaller than the sum of their radii. If so, an examination of overlaps of monomers between the chondrule and the dust particle is conducted, and a hit is detected if any pair of the monomers overlaps. At the same time, it is checked at each time step whether the two particles are moving away from each other, which indicates a missed collision. In the case of hit, it is further determined whether the incoming particle stick at the contact point, bounce or cause restructuring to the dust rim, by comparing the relative velocity at impact to the critical bouncing velocity and the critical rolling velocity (formula 2.6). 

 Upon a successful collision, the number of dust particles in each bin is adjusted according to the outcome, and the corresponding $ C_{ch,d}$ is updated based on the new equivalent radius of the chondrule and the change in the population of the dust particles, while in the case of a missed collision, the code proceeds to the next iteration and a new dust particle is selected. For each hit, miss, or restructuring event, the mass, radius, equivalent radius, relative velocity of the incoming particle, and the time interval between two interactions are recorded.
\\

\subsection{Numerical simulations}

We modeled dust accretion by chondrules at the midplane of a minimum-mass solar nebula (Hayashi 1981), at a distance of 1 AU from the sun, with a temperature of 280 K. The average molar mass, sound speed and molecular viscosity of the gas are 2.33 g/mol, $1.179\times 10^{5}$ cm/s and $1.8\times 10^{-4}$ g/(cm$\cdot$ s), respectively.

%The turbulent gas velocities $v_g$ in Eq. (\ref{eq:vturb}) were computed from an ideal magnetohydrodynamic (MHD) simulation of a protoplanetary disk in the shearing box approximation (Hawley et al. 1995). The turbulence was generated by the magnetorotational instability (MRI; Balbus \& Hawley 1998). Although mounting numerical evidence suggests that non-ideal MHD effects, such as ambipolar diffusion and Hall resistivity, play a crucial role in the overall magnetic evolution of PPDs and produce weak or no turbulence (Bai \& Stone 2013, Bai 2014, Simon et al. 2015), our MHD simulation provides upper limits on the possible values of $v_g$ that, in turn, give us upper bounds on the values of the turbulent relative velocities $v_T$ between dust and chondrules.

%In our simulations, the plasma environment is assumed to be hydrogen with electron and ion temperatures $T_e = T_i = 280$ K, as the plasma thermalizes with the gas due to collisions. In the case of low dust density, a negligible percentage of the electrons reside on the dust grains, and the number density of electrons and ions in the gas is set to be $n_{e}=n_{i}=3.5\times 10^{2}$ cm$^{-3}$ (Matthews et al. 2012). For high dust density, the ratio of free electrons to free ions is reduced due to the electron depletion, and we use $n_{e}/n_{i}=$ 0.1. We compare these two plasma environments with a neutral environment where the dust particles are not charged. 

The precise value of the turbulence strength in protoplanetary disks, as quantified by the $\alpha$ parameter, is uncertain, and the values often considered range from $\sim 10^{-6}$ to 0.1 (Hartmann et al. 1998; Cuzzi 2004; Ormel et al. 2008, Carballido 2011). In this study, we investigate $\alpha$ values of $10^{-1}, 10^{-2}, 10^{-3} , 10^{-4}, 10^{-5}, 10^{-6} $.

%The values of the parameters (chondrule radius, turbulence strength and electro-to-ion ratio) used in the simulations are listed in Table \ref{table1}. The \textit{first column} of Table \ref{table1} lists the value of the ratio $n_e/n_i$; the \textit{second column} records the value of the turbulent parameter $\alpha$; the \textit{third column} contains the value of the chondrule radius employed, and the \textit{fourth column} indicates the type of the aggregate: ``s'' represents sphere and ``agg'' represents aggregate. 

In the following, we adopt the form ``a$k$-r$\ell$'' for the simulation name,  in which ``a$k$" denotes the value of the turbulence strength $\alpha=10^{-k}$, and ``r$\ell$" refers to the chondrule radius, with $\ell$ specifying hundreds of microns. The labels suffixed by ``agg'' are simulations in which the dust particles are small aggregates consisting of up to $N$ spherical monomers. Otherwise, the dust particles are spheres. A description of the dust clusters is given in Sec. \ref{sec:effpaca}. We performed 10 runs for each simulation, and averaged the data for analysis.

%%%%%%%%%%%%%%%%%%%%%%%%%%%%%%%%%%%%%%%%%%%%%%%%%%%%%%%%%%%%%
\section{RESULTS}

%%%%%%%%%%%%%%%%%%%%%%%%%%%%%%%%%%%%%%%%%%%%%%%%%%%%%%%%%%%%%
\renewcommand{\theequation}{3.\arabic{equation}} \setcounter{equation}{0}
In order to analyze the structure and composition of a dust rim, we divided the rim into 30 horizontal layers (i.e., parallel to the chondrule’s surface), with each layer having the same thickness. We define the porosity of each layer as the ratio of the volume of voids within the layer, which is the total volume of the layer minus the sum of the volume of the monomers (or monomer portions) within that layer, to the total volume of the layer. To avoid edge effects of the dust pile, only the inner region of the pile with a radius of 50$\%$ of the total pile radius (i.e., the distance from the pile center to pile edge) is analyzed.

%\begin{figure*}
%\includegraphics[width=9cm]{rr.pdf}
%\caption{Illustration of radius and equivalent radius for one aggregate. The black circle indicates the radius $R$, defined as the the maximum distance from the center of mass to constituent monomers. The red circle indicates the equivalent radius $R _{\sigma}$, as defined in the text. }
%\label{fig1}
%\end{figure*}

\subsection{Effect of turbulence strength}\label{sec:effturb}

\begin{figure*}
\includegraphics[width=9cm]{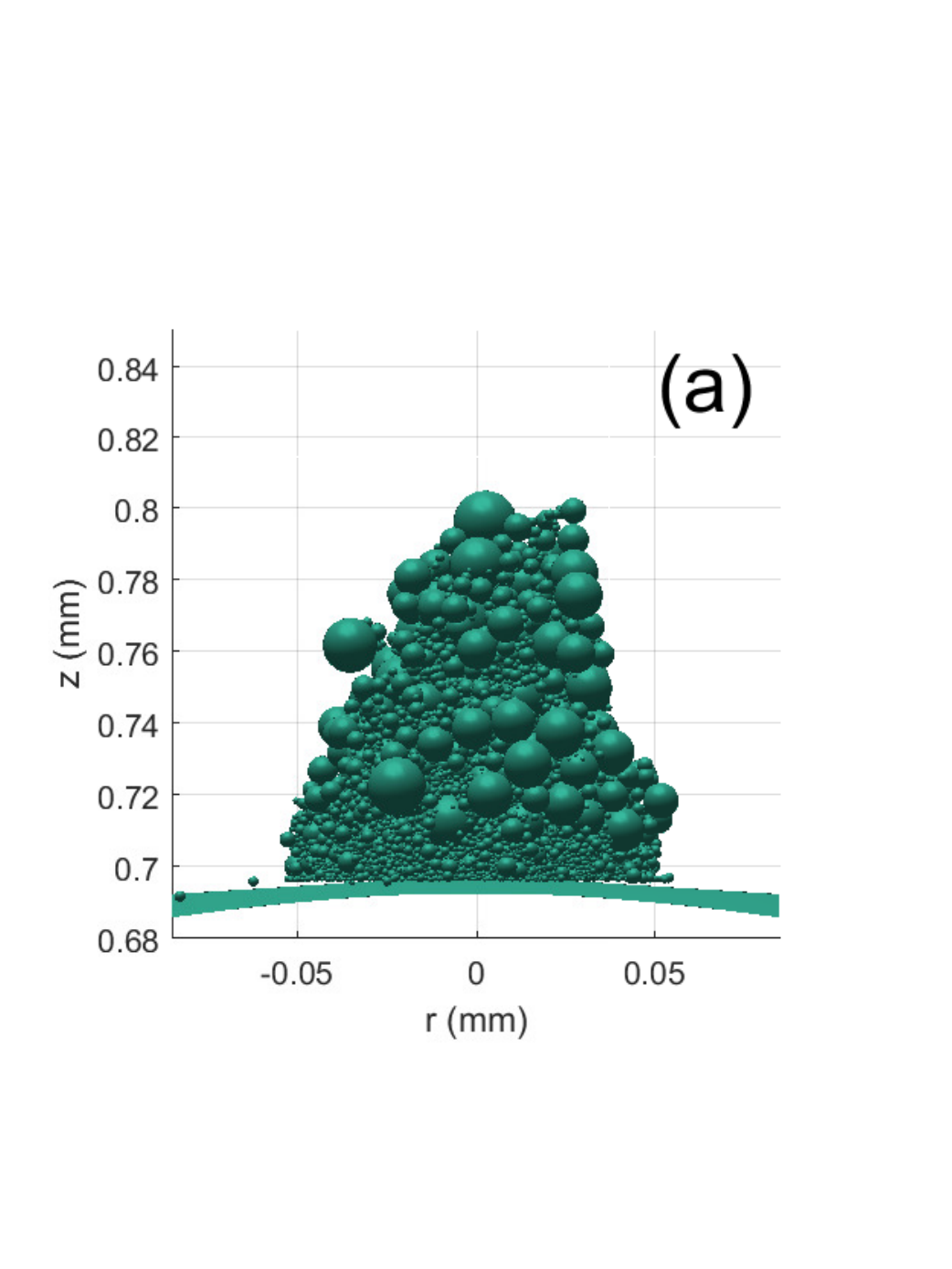}\includegraphics[width=9cm]{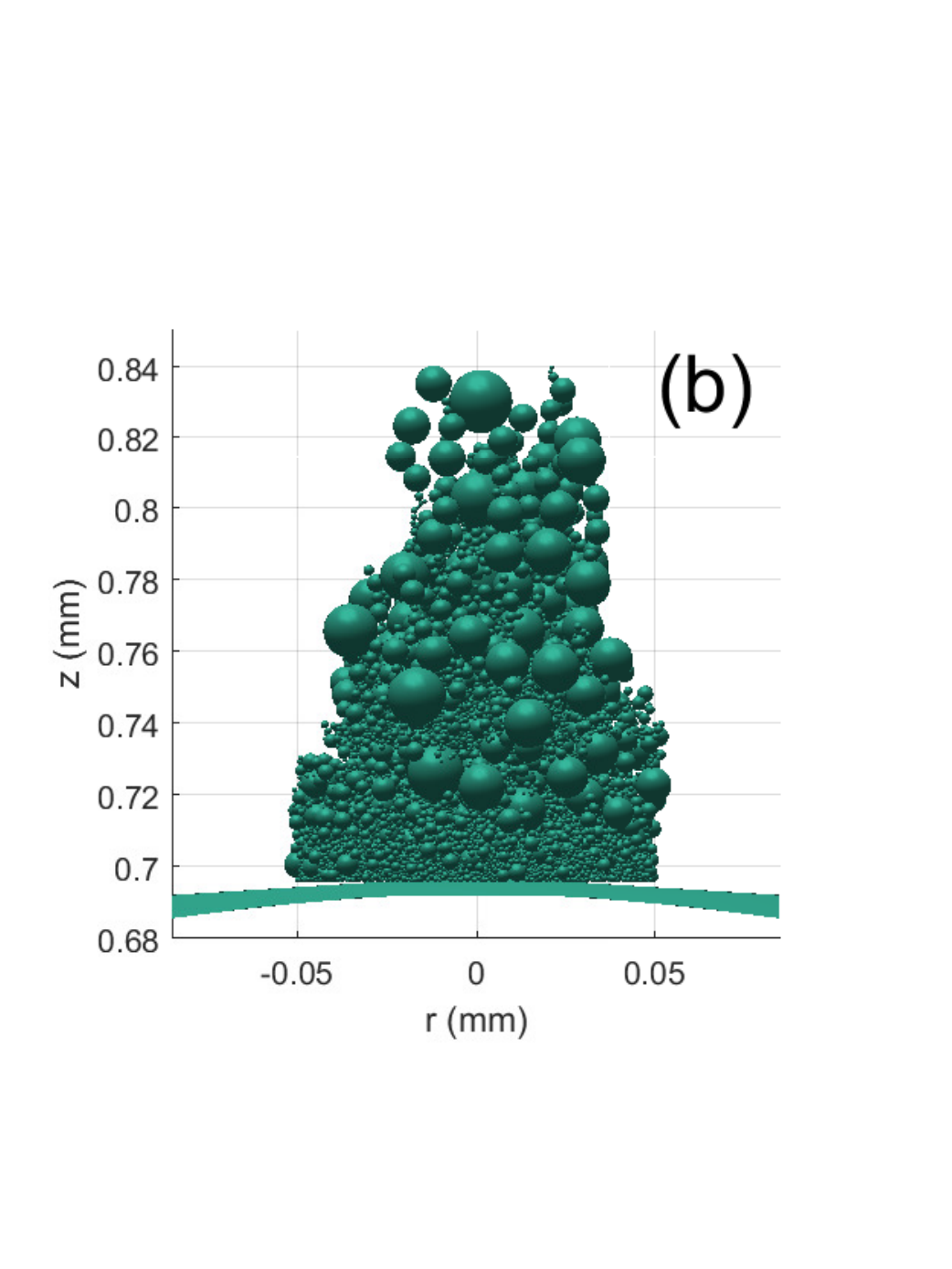}
\includegraphics[width=9cm]{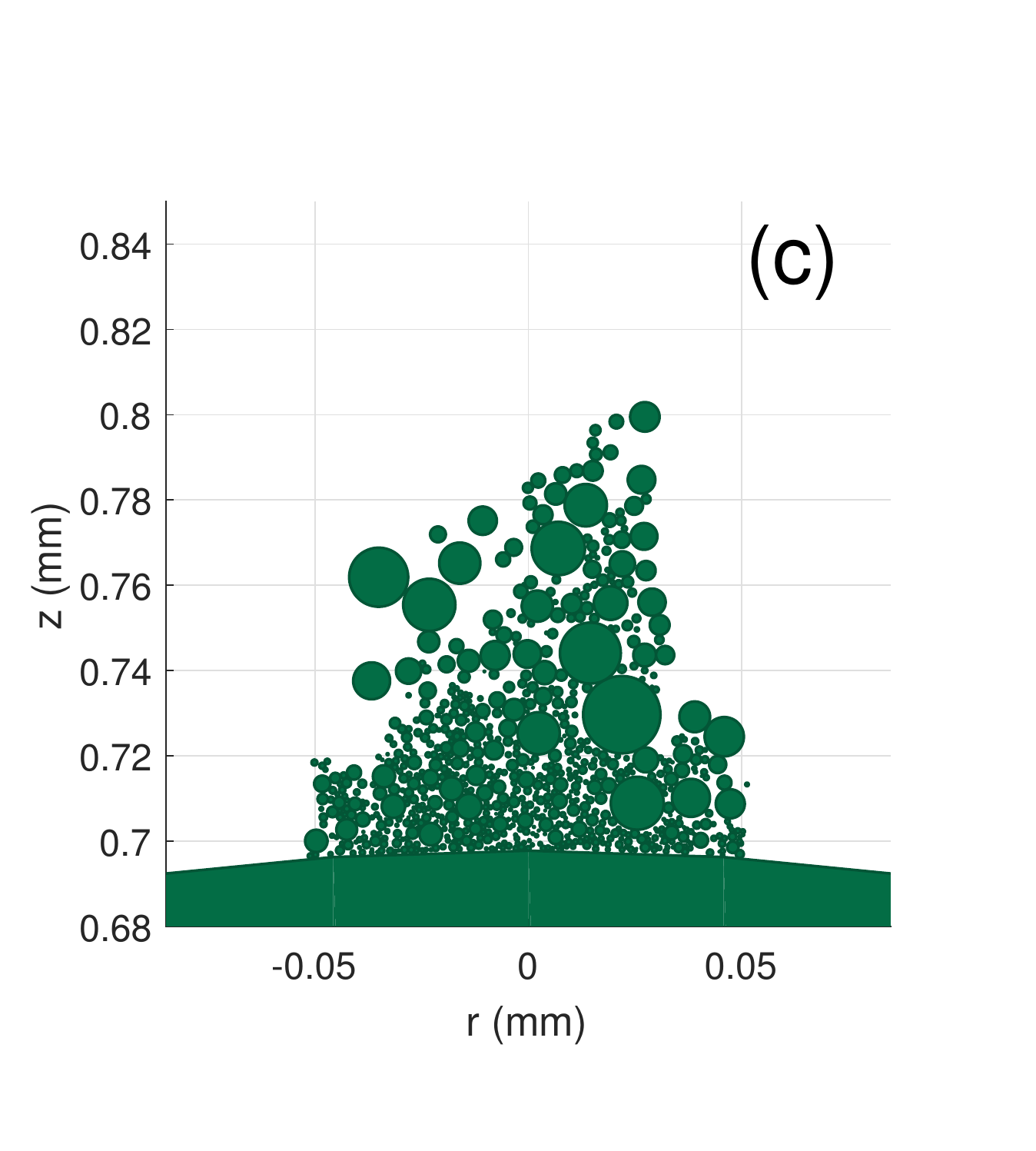}\includegraphics[width=9cm]{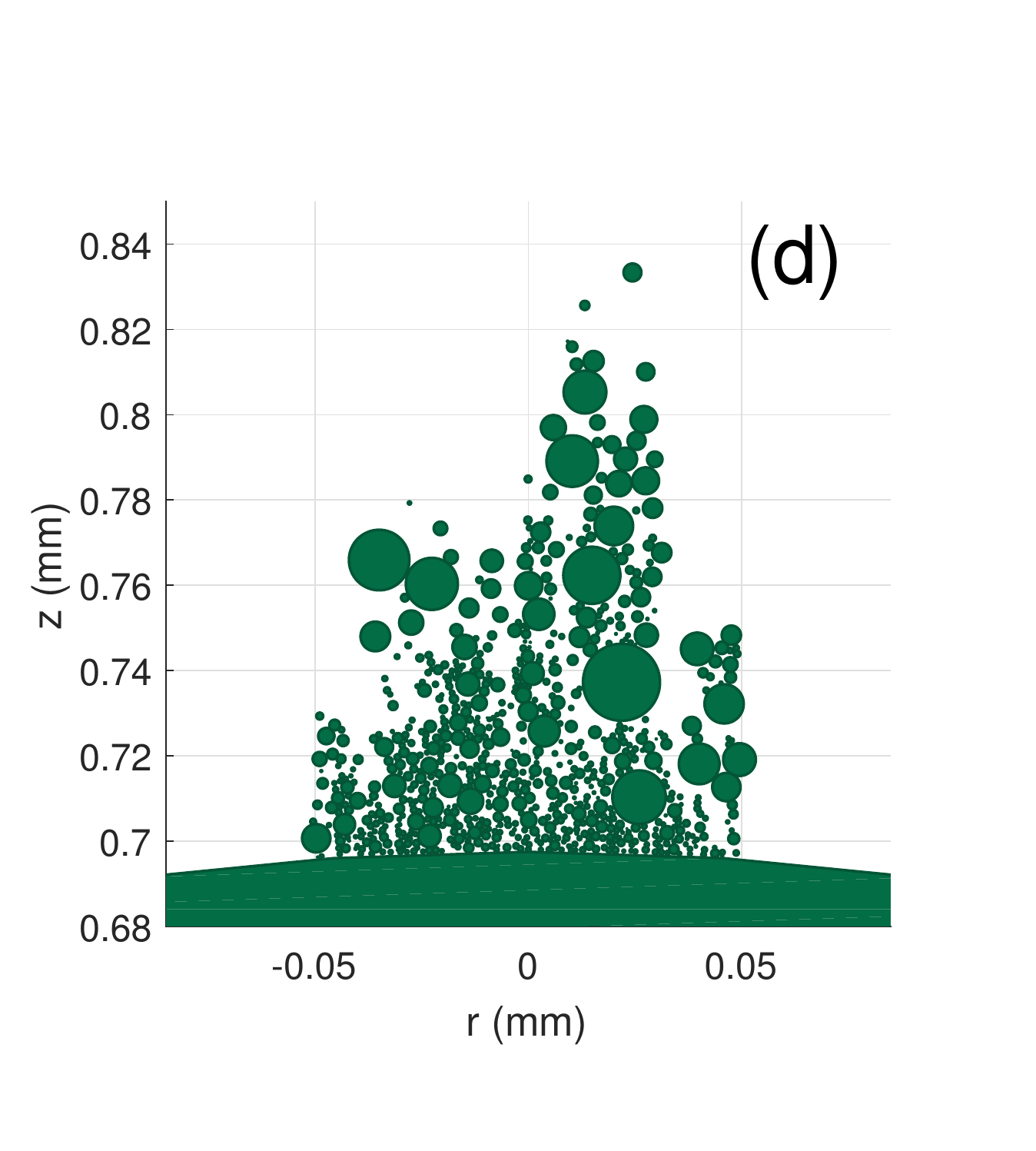}
\caption{Rim growth on a 100 $\mu$m-diameter patch on the surface of a chondrule with a radius of 700 $\mu$m, from runs a1-r7 (a,c) and a6-r7 (b,d). The elapsed times are 0.26 yr and 173.61 yr, respectively. Panels (c) and (d) show the monomer distributions on a vertical plane cutting through the center of the dust piles that are shown in (a) and (b), respectively. More compaction (less open spaces) is observed in the strong-turbulence case (c). The apparent detachment of some monomers and groups of monomers from the main dust pile is due to the slicing procedure, which cuts off other monomer structures that support the ``detached" ones. }
\label{fig:f1}
\end{figure*}

Figure \ref{fig:f1} shows dust piles on the surface of a chondrule with a radius of 700 $\mu$m, for runs a1-r7 (Fig. \ref{fig:f1}a) and a6-r7 (Fig. \ref{fig:f1}b). The piles are formed on circular patches of 100 $\mu$m in diameter. The respective turbulent strengths are $\alpha=10^{-1}$ and $10^{-6}$. The piles contain 30,000 monomers. Since the relative speeds between chondrules and dust particles are higher in regions of stronger turbulence, massive particles are more likely to cause restructuring, as they have a larger kinetic energy which can exceed the threshold rolling energy. Therefore, chondrule dust rims formed in environments with stronger turbulence are more compact than those formed in weak turbulence. This can be better appreciated in Fig. \ref{fig:f1}c and Fig. \ref{fig:f1}d, which show a vertical slice through the center of the dust pile of Figs. \ref{fig:f1}a and \ref{fig:f1}b, respectively. There are more open spaces between monomers in the weak-turbulence case (Fig. \ref{fig:f1}d), a sign of less compaction (note that monomers and groups of monomers that are apparently detached from the main pile in Figs. \ref{fig:f1}c and \ref{fig:f1}d appear to be so due to the slicing procedure, which cuts off other monomer structures on which the ``detached" structures are supported).

\begin{figure*}[!htb]
\includegraphics[width=9cm]{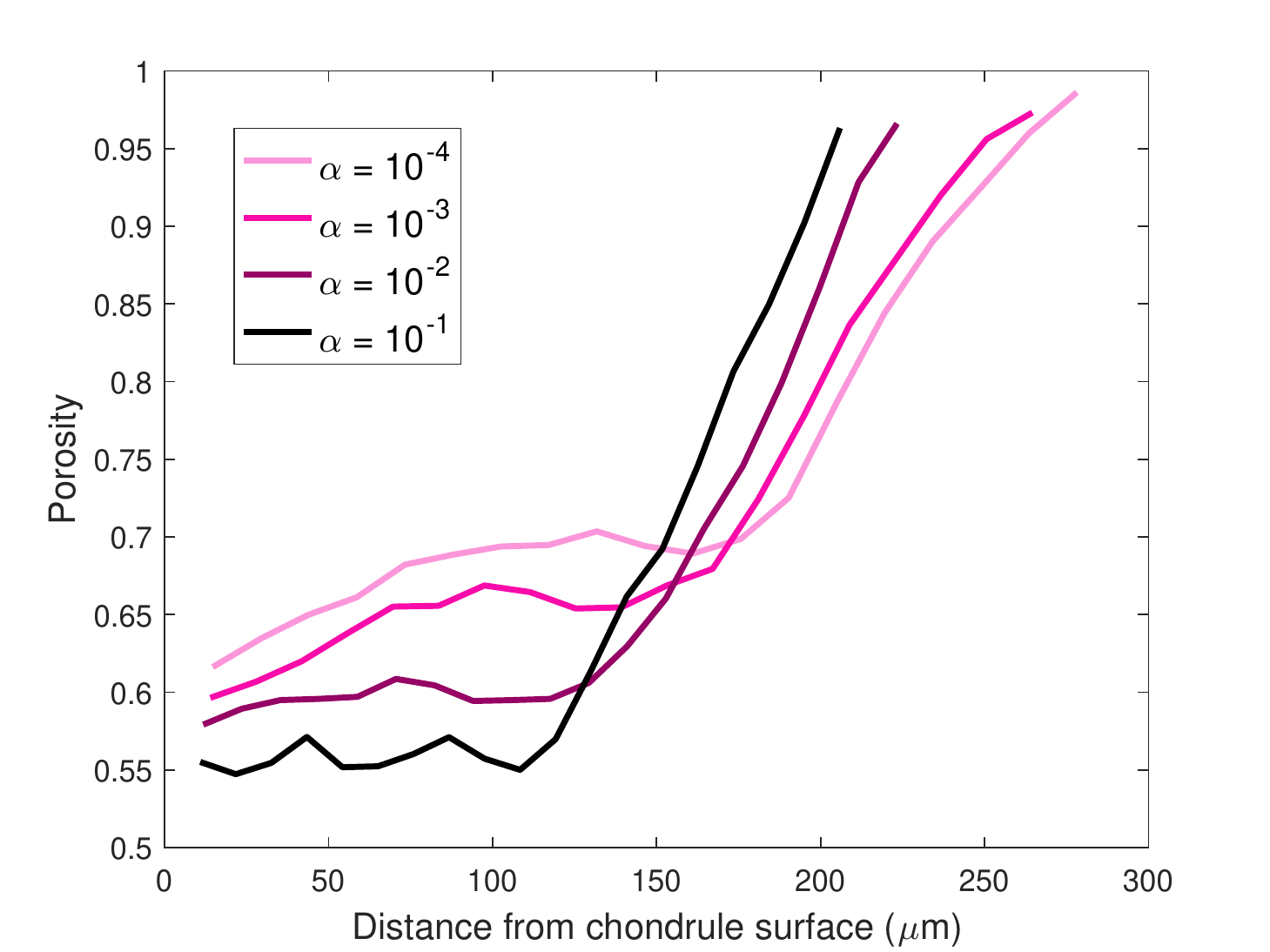}
\caption{Radial profiles of dust rim porosity in each horizontal layer on a chondrule with a 600-$\mu$m radius, for four different turbulent strengths ($\alpha =10^{-k}$, corresponding to runs a$k$-r6, $k=1,2,3,4$.), after accretion of $N$ = 60,000 monomers, for particle aggregation (PA). The elapsed times are 0.52 yr (for $\alpha=10^{-1}$), 1.62 yr ($\alpha=10^{-2}$), 5.04 yr ($\alpha=10^{-3}$) and 14.05 yr ($\alpha=10^{-4}$). The position of the ``knee" in the porosity curve is defined as the rim thickness.}
\label{fig:f3}
\end{figure*}

Figure \ref{fig:f3} shows the rim porosity as a function of distance from the chondrule surface, for different turbulence strengths. Note that the porosity increases rapidly at a certain radial distance in the dust pile. Thus we define the rim thickness by the position of the ``knee" in the porosity plot. The rims formed in weak turbulence ($\alpha=10^{-4}$) are more porous than those formed in strong turbulence ($\alpha=10^{-1}$) after accretion of 60,000 monomers. Strong turbulence results in a more compact rim, and restructuring leads to an approximately constant porosity of 55\% throughout the thickness of the rim. Weak turbulence produces less restructuring, resulting in rims which are overall more porous, and the porosity increases from the base of the rim to the top, as very fine dust particles are able to pass through voids and fill in the lower rim layers. Analysis of chondrules with radii between 500 and 1000 $\mu$m shows that FGRs formed in weak turbulence ($\alpha\leq10^{-4}$) have an average porosity of 60-74\%, while those formed in strong turbulence ($\alpha=10^{-1}$) have an average porosity of 52-60\%, with the lower end of each range corresponding to large chondrules and the upper end to small chondrules. In general, the porosity of FGRs decreases with both turbulence (stronger dependence) and chondrule radius (weaker dependence). The effect of chondrule size will be further discussed in section \ref{sec:effch}.

%Figure \ref{f3}b shows that the dust rims formed in high turbulence have slightly greater portion of small monomers throughout the thickness than those formed in weak turbulence, and the upper layers tend to have a higher ratio of large monomers to small monomers than the lower layers. The cases of $\alpha =10^{-5}, 10^{-6}$ are not included, since they are quite similar to the case of $\alpha =10^{-4}$, for which turbulence is low enough that almost no restructuring takes place.

\begin{figure*}[!htb]
\includegraphics[width=9cm]{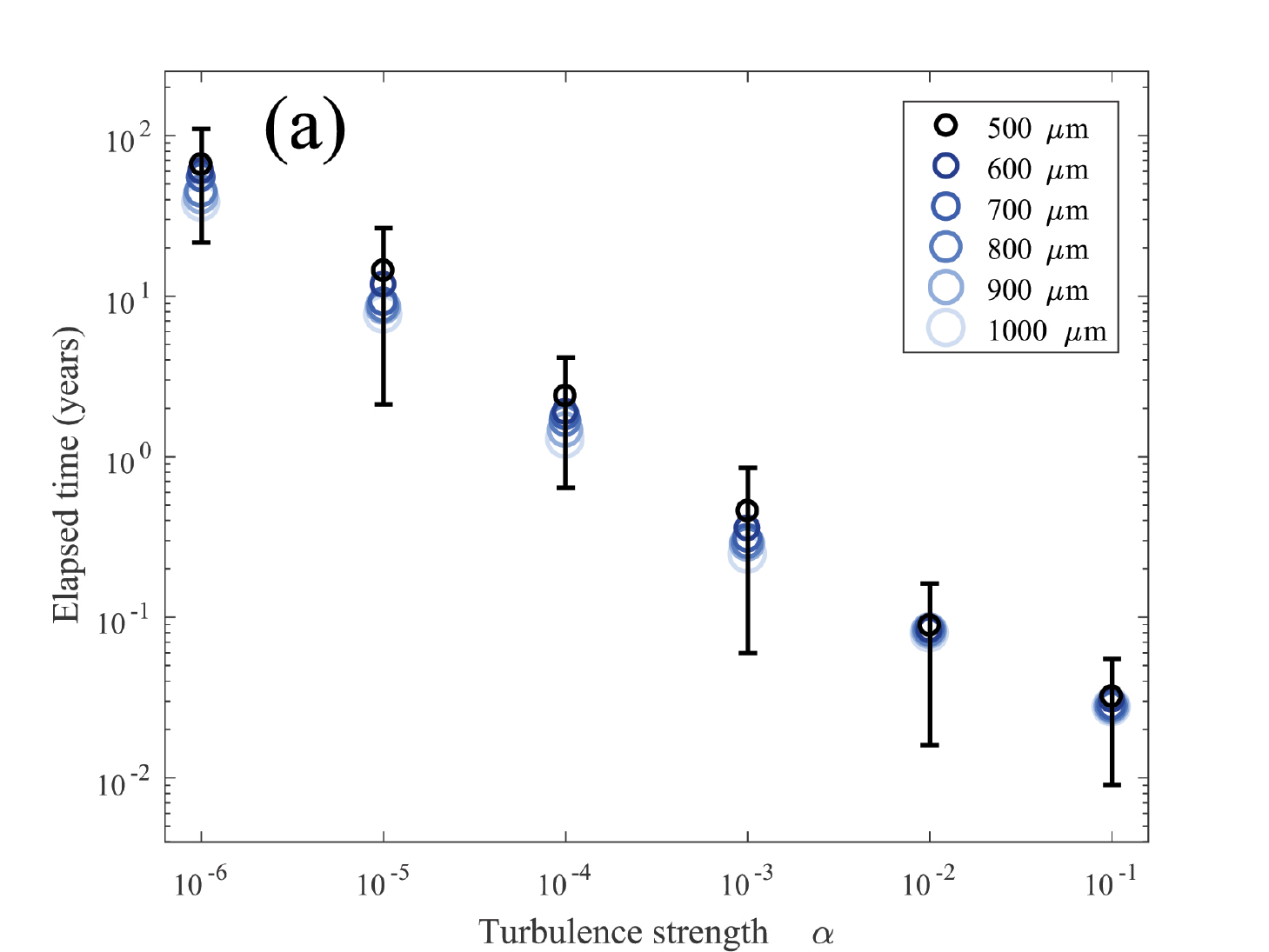}\includegraphics[width=9cm]{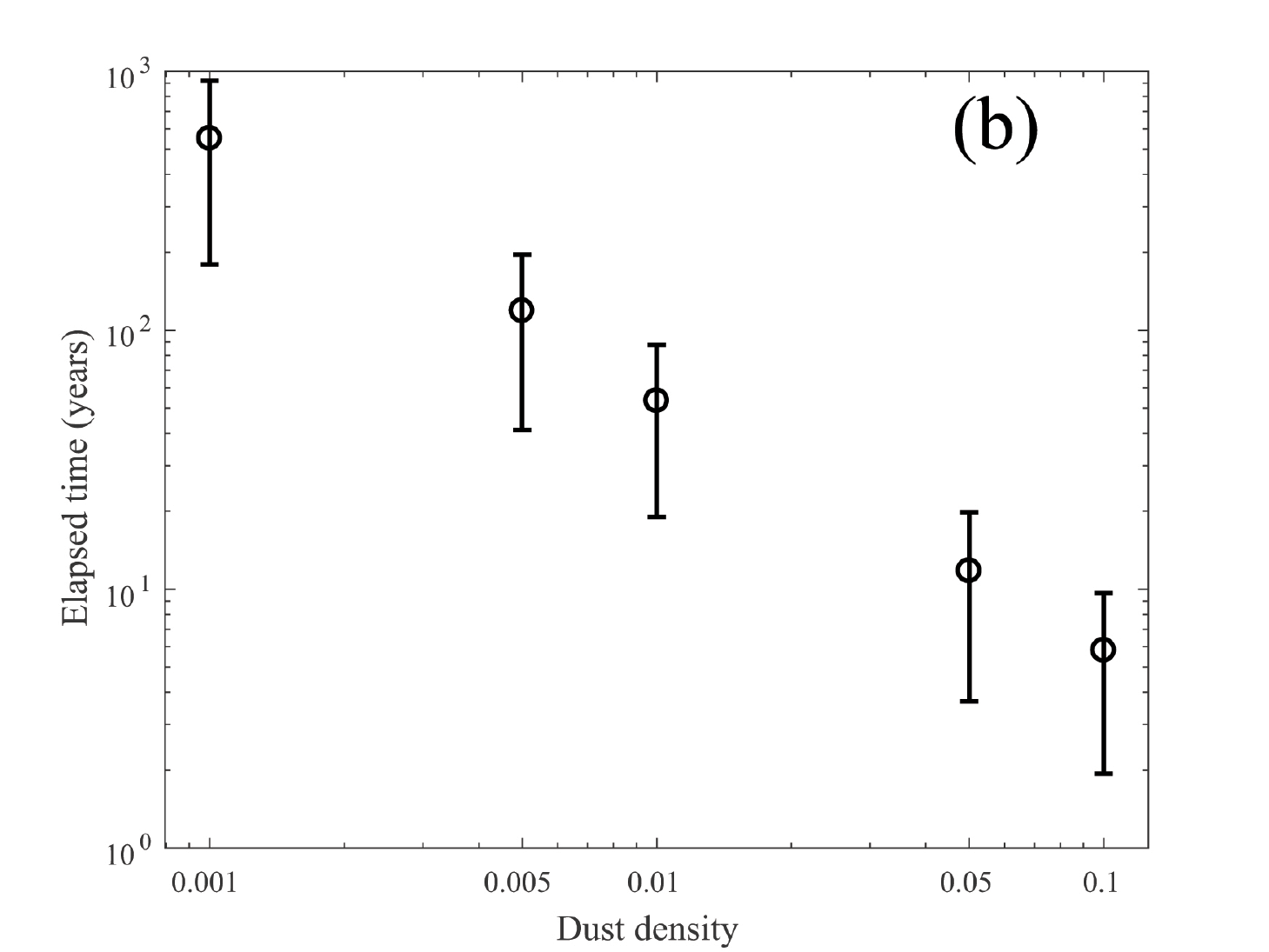}
\caption{a) Elapsed time to build FGRs for different turbulence strengths and different chondrule radii (runs a$k$-r$\ell$, $k=1,2,...6$, $\ell=5,6,...,10$). The midpoints correspond to a thickness of 180 $\mu$m for each chondrule size, while the lower and upper bounds (shown for the 500-$\mu$m-radius chondrule) indicate the times to build a thickness of 40 $\mu$m and 320 $\mu$m, respectively. b) Elapsed time to build FGRs on a 500-$\mu$m-radius chondrule for different dust densities with $\alpha =10^{-6}$. The meaning of the upper and lower bounds as well as the midpoints is the same as (a). }
\label{fig:f3_2}
\end{figure*}

In addition to the enhanced restructuring, the increased relative velocity caused by stronger turbulence also leads to more frequent collisions between the chondrule and the dust particles, speeding up the growth of the dust rim. Figure \ref{fig:f3_2}a displays the time necessary to build rims of three different thicknesses on a chondrule of 500 $\mu$m in radius, in each of the six turbulence conditions that we consider. The lower bound of each vertical bar represents a rim thickness of 40 $\mu$m, while the upper bound corresponds to a thickness of 320 $\mu$m. The midpoints represent a thickness of 180 $\mu$m. The time required to build rims of 180 $\mu$m in thickness is 2.7-3.2$\times 10^{-2}$ yrs for $\alpha = 10^{-1}$; 7.9-8.9$\times 10^{-2}$ yrs for $\alpha = 10^{-2}$; 2.4-4.6$\times 10^{-1}$ yrs for $\alpha=10^{-3}$; 1.3-2.4 yrs for $\alpha=10^{-4}$; 7.6-14 yrs for $\alpha=10^{-5}$; and 38-66 yrs for $\alpha =10^{-6}$ (the lower end of each range corresponds to large chondrules and the upper end to small chondrules; see section \ref{sec:effch} for more detailed discussion). As expected, the time needed to build a rim of a given thickness decreases with increasing turbulence strength. 

Using a linear fit to relate FGR thickness to elapsed growth time, we obtain the following growth rates (the increase in the rim thickness per unit time): 6100 $\mu$m/yr for $\alpha = 10^{-1}$; 1900 $\mu$m/yr for $\alpha = 10^{-2}$; 350 $\mu$m/yr for $\alpha=10^{-3}$; 80 $\mu$m/yr for $\alpha=10^{-4}$; 10 $\mu$m/yr for $\alpha=10^{-5}$; and 3 $\mu$m/yr for $\alpha =10^{-6}$. We also investigated the effect of the dust density on the growth rate. Figure \ref{fig:f3_2}b shows that the higher dust density leads to lower elapsed time for the same turbulence condition ($\alpha =10^{-6}$), since a larger dust population can cause more frequent collisions between the chondrules and dust particles. The growth rates are 0.4, 2, 4, 17 and 36 $\mu$m/yr for the ratios of dust density to gas density equal to 0.001, 0.005, 0.01, 0.05 and 0.1, respectively.

%\begin{figure*}[!htb]
%\includegraphics[width=9cm]{f51_2.eps}\includegraphics[width=9cm]{f52_2.eps}
%\caption{Elapsed time to build FGRs of three different thicknesses (represented by the upper and lower bounds of the vertical bars, as well as by the points) for different turbulence strengths (runs a$k$-r5, $k=1,2,...6$) and b) different dust densities for $\alpha =10^{-6}$, on a 500-$\mu$m-radius chondrule. The upper and lower bounds of each bar correspond to FGR thicknesses of 320 $\mu$m and 40 $\mu$m respectively, while the points correspond to a thickness of 180 $\mu$m. Points of other chondrule radii (600-1000$\mu$m) are plotted in (a) for comparison.}
%\label{fig:f3_2}
%\end{figure*}

\subsection{Effect of chondrule size}\label{sec:effch}

\begin{figure*}[!htb]
\includegraphics[width=9cm]{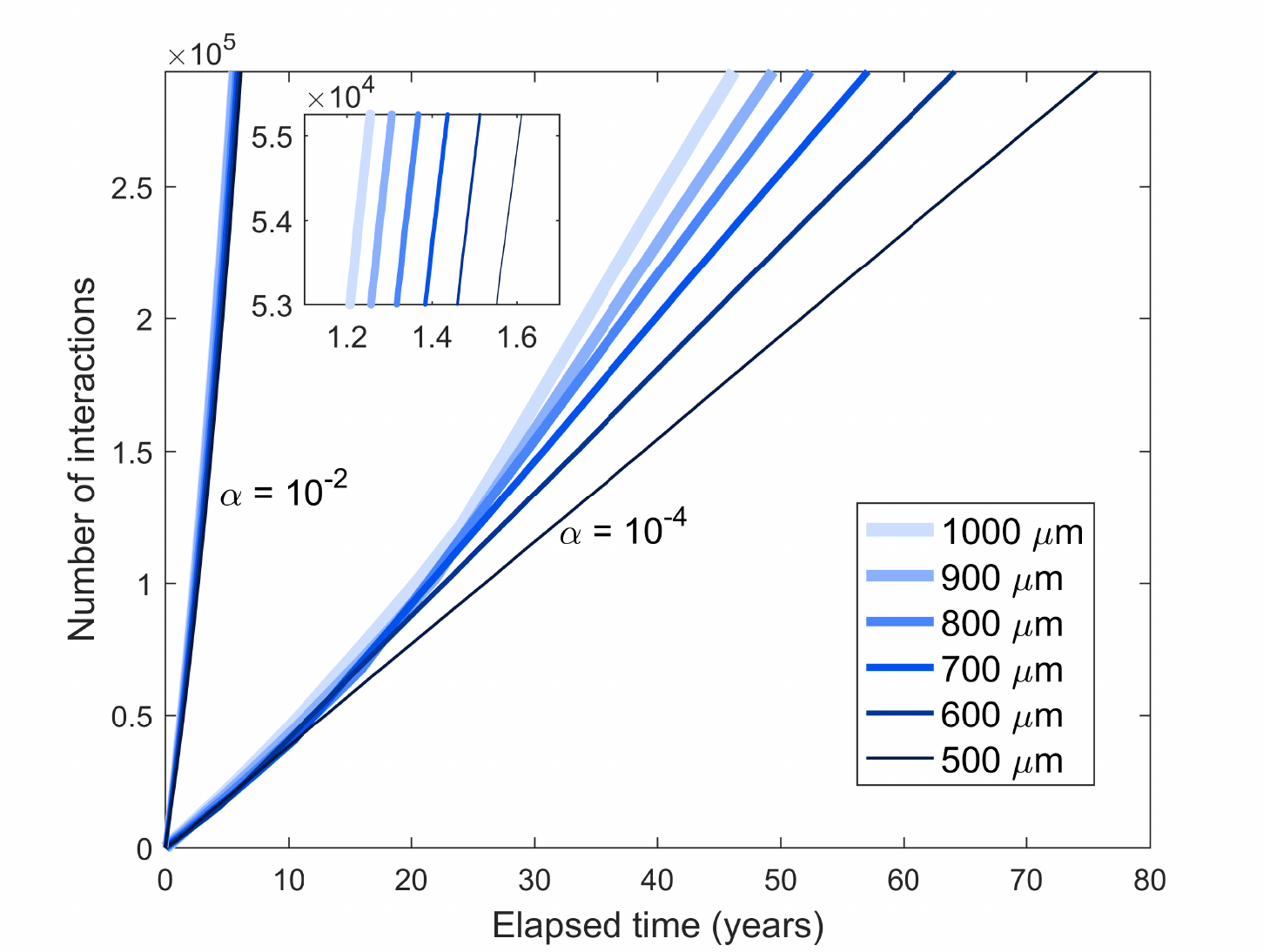}
\caption{Number of interactions as a function of elapsed time for different chondrule radii, with $\alpha =10^{-4}$ (runs a4-r$\ell$, $\ell=5,6,...,10$) and $10^{-2}$ (runs a2-r$\ell$, $\ell=5,6,...10$). The inset is a magnification of the $\alpha =10^{-2}$ curves between 1.1 and 1.7 years.}
\label{f5}
\end{figure*}

Another important factor affecting dust rim growth is the size of the chondrule. Larger chondrules have higher relative velocity with respect to the dust particles [Eq. (\ref{eq:vturb})], which, together with their greater cross-sectional area, increase the collision rate between the chondrules and the dust particles. Figure \ref{f5} shows the number of interactions as a function of the elapsed time for different chondrule sizes and turbulence strengths. The small chondrules in weak turbulence ($\alpha =10^{-4}$) take the longest time to collide with the dust particles. The broader separation between the curves in weak turbulence shows that the chondrule size has a greater impact on the collision rate when the relative velocity is low. 

\begin{figure*}[!htb]
\includegraphics[width=9cm]{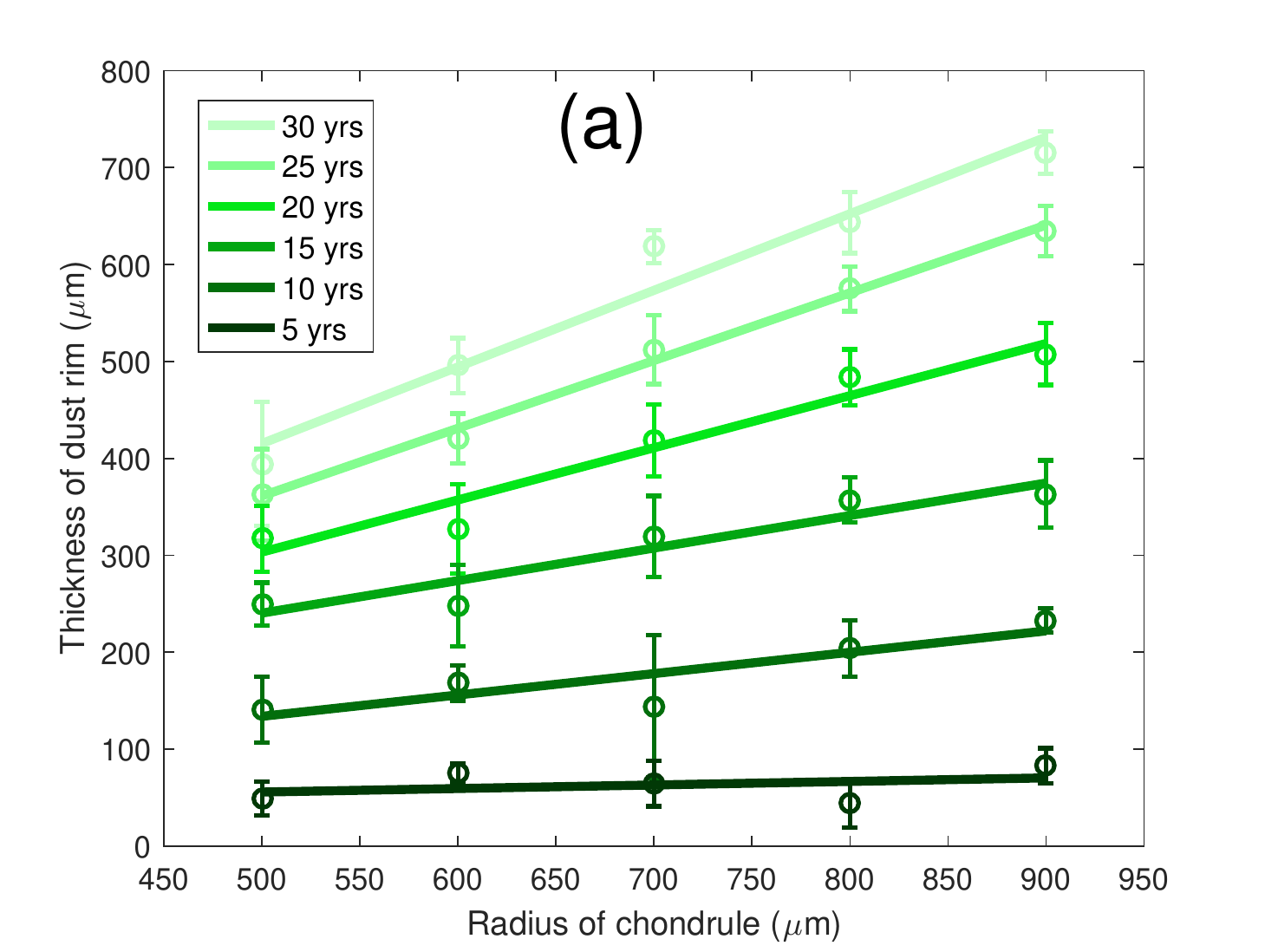}\includegraphics[width=9cm]{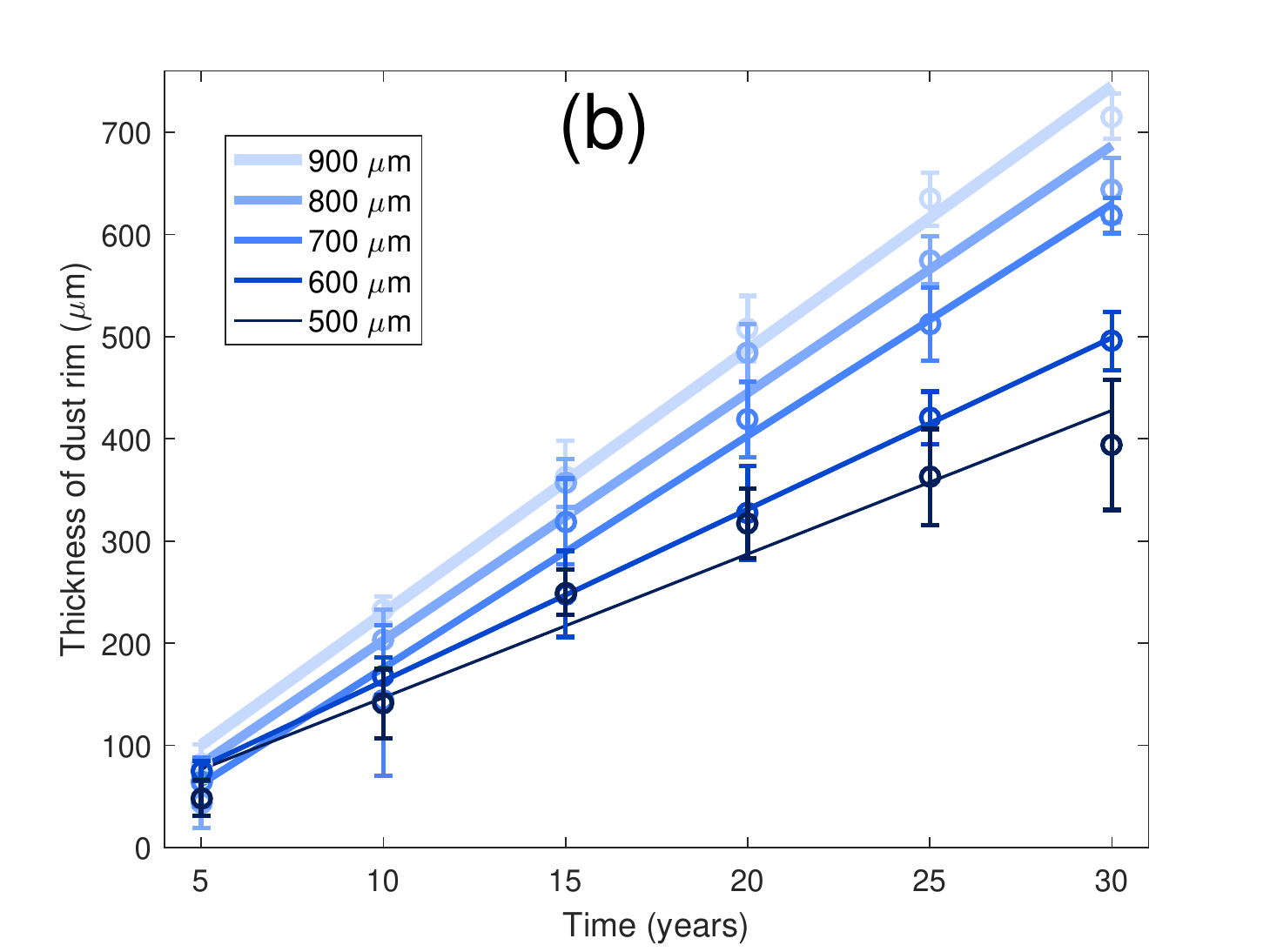}
\caption{a) Thickness of dust rims formed through the addition of single monomers (PA) after $t=5, 10, 15, 20, 25, 30$ years, with a turbulent strength $\alpha =10^{-5}$ and for different chondrule radii. The data points correspond to runs a5-r$\ell$, $\ell=5, 6,..., 9$. The lines are linear, least-square polynomial fits to the data points. b) Same as a), except that the thickness of dust rim is a function of the elapsed time, and each line represents a chondrule size.}
\label{f6}
\end{figure*}

In spite of the higher collision rate, more dust is required to build a rim of a certain thickness for large chondrules than for smaller chondrules. As a consequence of these two factors, the rim thickness scales linearly with the chondrule radius, as shown in Fig. \ref{f6}a, consistent with measurements by Paque \& Cuzzi (1997), Hanna and Ketcham (2018), and simulations by Ormel et al. (2008) and Carballido (2011). The linear relationship becomes steeper over time, as the greater collision cross section and larger relative velocities of the larger chondrules increases the collision rate, causing large chondrules to grow even faster. The growth rates of chondrules of different sizes are also shown in Fig. \ref{f6}b, in which the slopes of the lines increase with the chondrule size. 

\begin{figure*}[!htb]
\includegraphics[width=9cm]{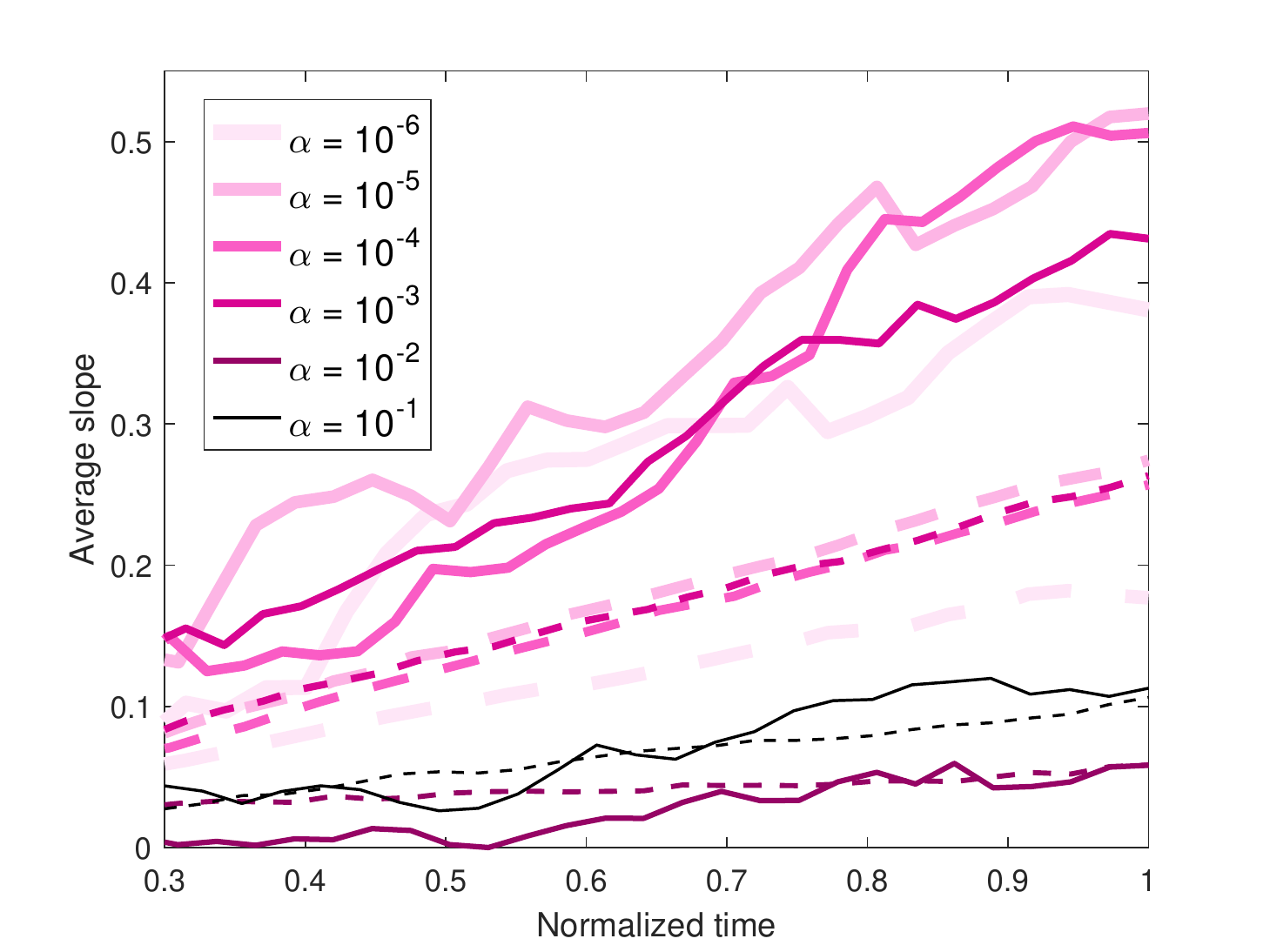}
\caption{Time evolution of the slopes of rim thickness versus chondrule size (solid lines), with the elapsed time for each turbulence strength normalized by its maximum time (the time required to build a dust rim with a thickness of 350 $\mu$m). The dashed lines are for an ideal case where the dust rims are fully compacted (with zero porosity), indicating the influence of the porosity on the slope.}
\label{slope}
\end{figure*}

The comparison of the time evolution of the slopes for different turbulence strengths is shown in Fig. \ref{slope}. All the rims have a thickness of 350 $\mu$m at the maximum elapsed times, which are  0.05 yr (for $\alpha=10^{-1}$), 0.15 yr ($\alpha=10^{-2}$), 0.66 yr ($\alpha=10^{-3}$), 3.37 yr ($\alpha=10^{-4}$), 20.98 yr ($\alpha=10^{-5}$), and 92.89 yr ($\alpha=10^{-6}$); the elapsed time for each turbulence strength is normalized by its maximum time. Although the slope increases with time for all turbulence strengths, the slope is greater and increases faster in weak turbulence than in strong turbulence, meaning that the growth rates of dust rims are less sensitive to chondrule size in strong turbulence. The dashed lines in Fig. \ref{slope} indicate the minimum thickness of of the dust rim, if all of the constituent dust material were compressed to a solid layer of zero porosity on the surface.  This gives a lower bound of the rim thickness which may be observed in a chondrule after restructuring.

%\begin{figure*}[!htb]
%\includegraphics[width=9cm]{f7_4.eps}
%\caption{Thickness of FGRs as a function of chondrule radius for PA runs a$k$-r$\ell$, $k=2, 5, 6$, $\ell=5, 6,..., 9$. Data points are mean values, and error bars represent one standard deviation, for $\alpha =10^{-5}$. The solid lines are linear, least-square fits to the data points. The different linear fits correspond to different times. Starting at the bottom linear fit, times are $t=5, 10, 15, 20, 25$ and 30 years. Linear fits for the cases of $\alpha =10^{-2}$ (dottted lines) and $\alpha =10^{-6}$ (dashed lines) are plotted for comparison. Note that the maximum time is 2.1$\times 10^{-1}$ yrs for $\alpha =10^{-2}$ and 1.4$\times 10^{2}$ yrs for $\alpha =10^{-6}$, and the time interval between neighboring lines are the same in each case.    }
%\label{f6}
%\end{figure*}

\begin{figure*}[!htb]
\includegraphics[width=9cm]{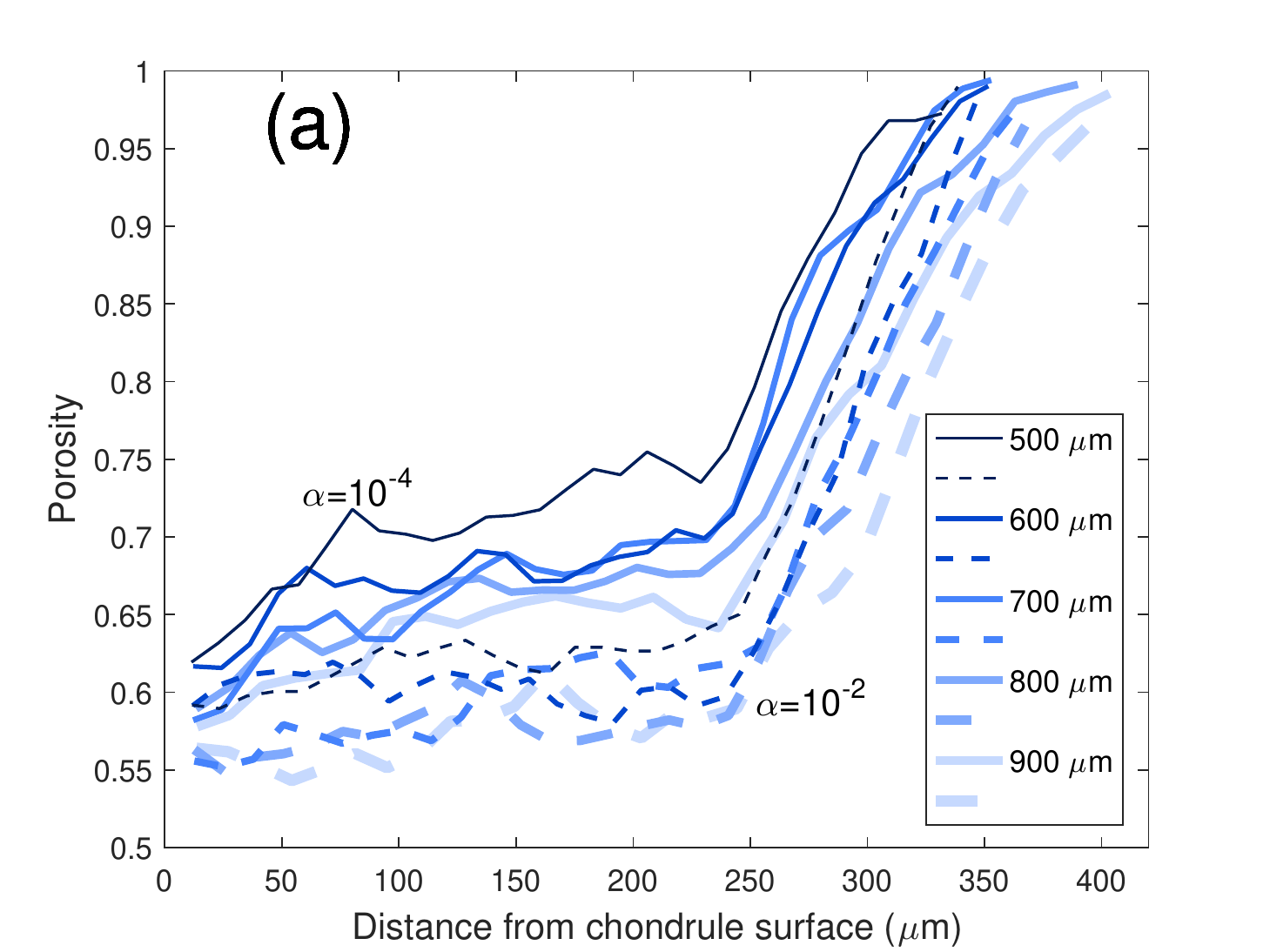}\includegraphics[width=9cm]{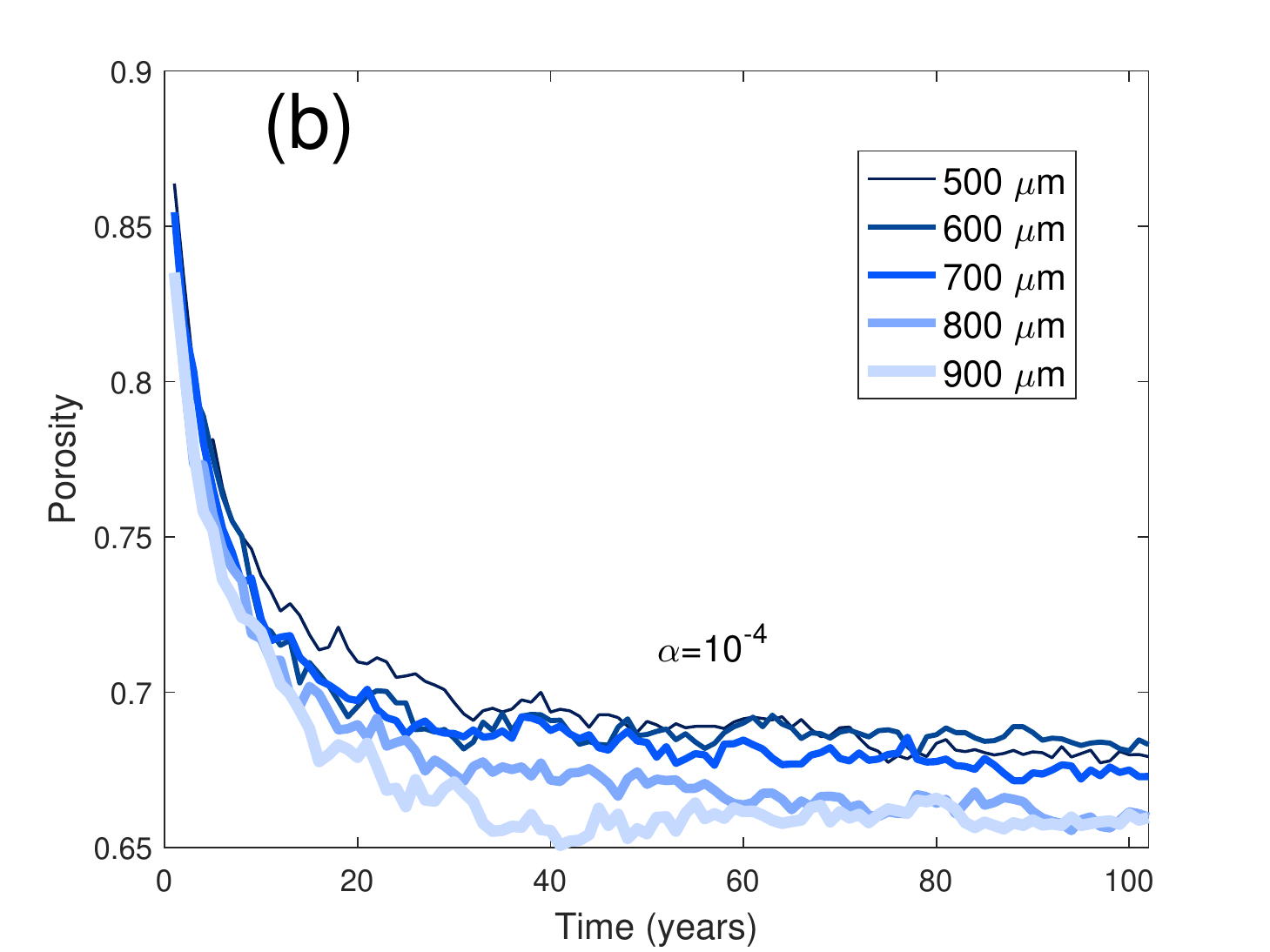}
\caption{a) Radial profiles of dust rim porosity on chondrules of different sizes, for turbulent strengths $\alpha =10^{-4}$ (\textit{solid curves}, corresponding to runs a4-r$\ell$, $\ell=5, 6, ..., 9$) and $\alpha=10^{-2}$ (\textit{dashed curves}, corresponding to runs a2-r$\ell$, $\ell=5, 6, ..., 9$), for equal rim thickness. The data were obtained by particle aggregation (PA). b) Mean porosity of the rim (the top-most layer with porosity $> 0.9$ is discarded; note that this value is higher than the cutoff porosity for calculating the rim thickness, and this is an overall porosity of almost the whole rim), formed through particle aggregation (PA), as a function of time and for different chondrule radii. The turbulent strength is $\alpha =10^{-4}$. The data correspond to runs a4-r$\ell$, $\ell=5, ... , 9$. }
\label{f7}
\end{figure*}

In addition to the change in the collision rate, the velocity difference caused by different chondrule sizes also affects the restructuring and therefore the porosity of the dust rim. Figure \ref{f7}a shows that larger chondrules are enveloped in more compact dust rims than smaller chondrules with the same rim thickness for both strong and weak turbulence ($\alpha =10^{-4}, 10^{-2}$), as larger chondrules have more kinetic energy for restructuring. Figure \ref{f7}b shows the time evolution of the overall porosity of the dust rims. The colliding dust particles constantly fill in the gaps of the existing rim and form a new outermost layer which has the highest porosity. The ratio of the volume of the outermost layer to the volume of the inner layer gets lower as the rim becomes thicker. This, together with the fact that larger chondrules experience more restructuring, results in the decrease in porosity of the whole rim over time, for all chondrule sizes, until a certain time has passed, after which the porosity is constant.  

\subsection{Comparison between PA and CA}\label{sec:effpaca}

\begin{figure*}[!htb]
\includegraphics[width=9cm]{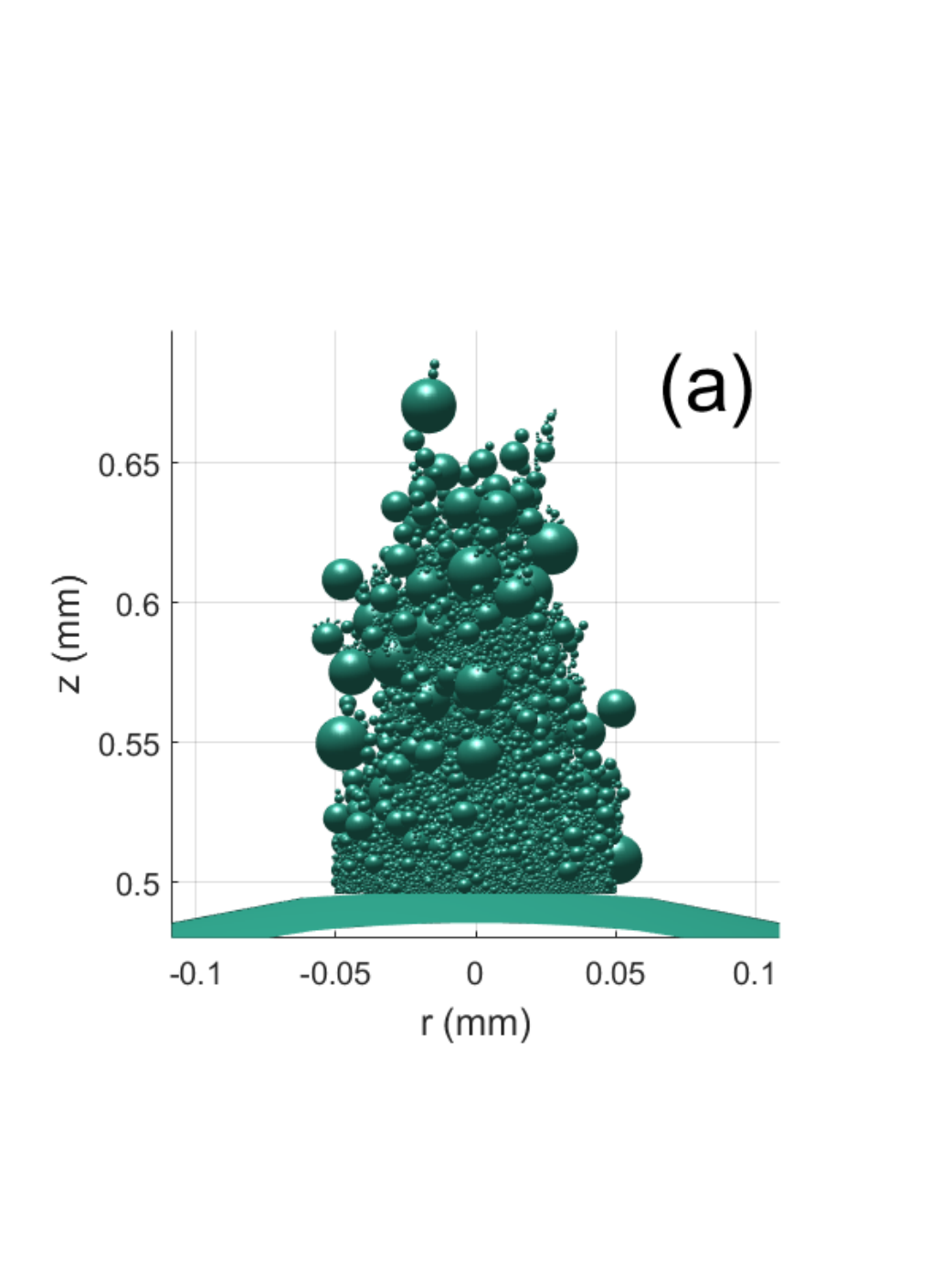}\includegraphics[width=9cm]{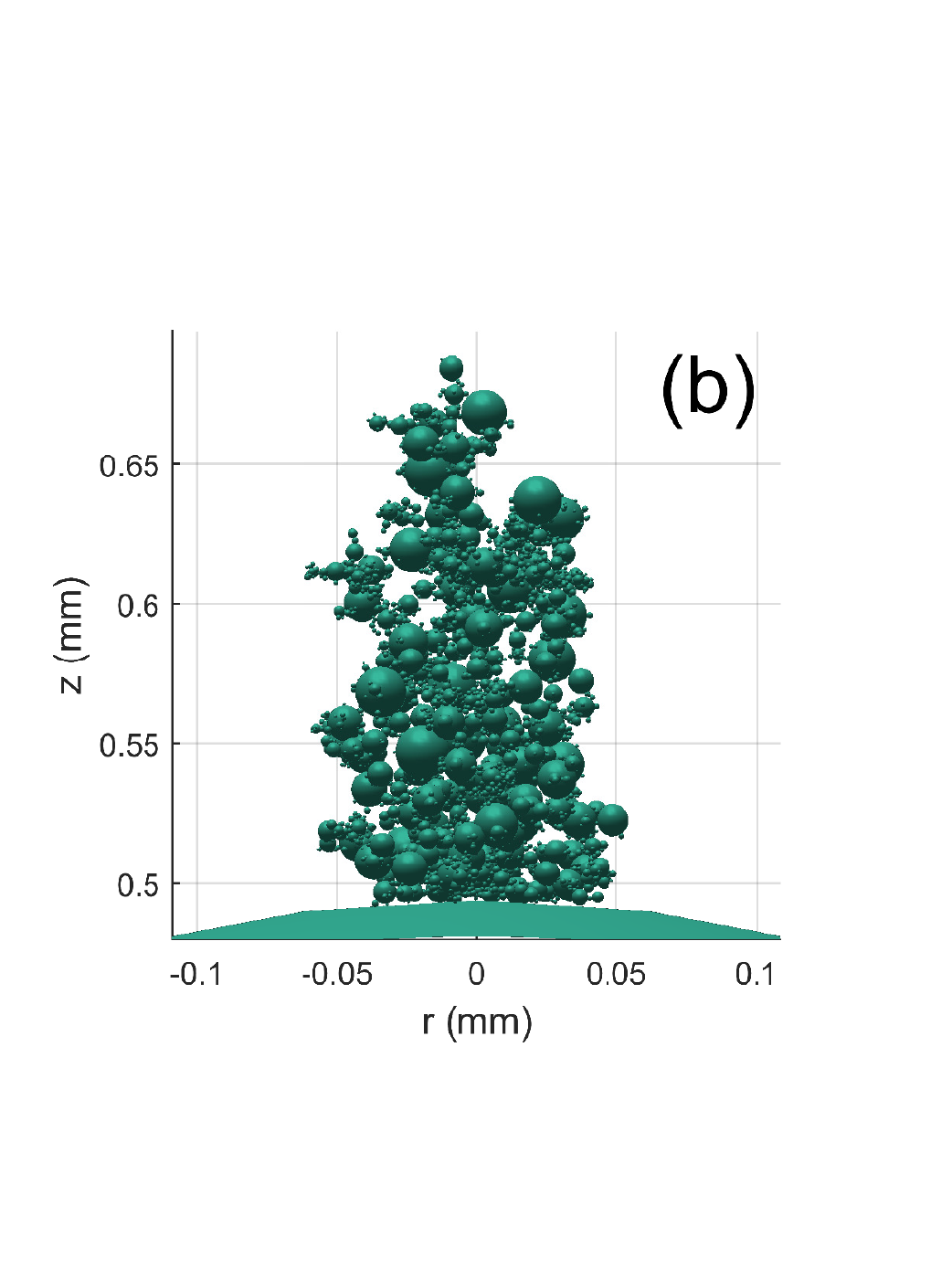}
\caption{Rim growth on a 100 $\mu$m-diameter patch on the surface of a chondrule with a radius of 500 $\mu$m, formed through a) particle aggregation (PA, run a4-r5) and b) cluster aggregation (CA, run a4-r5-agg).}
\label{f15}
\end{figure*}

\begin{figure*}[!htb]
\includegraphics[width=9cm]{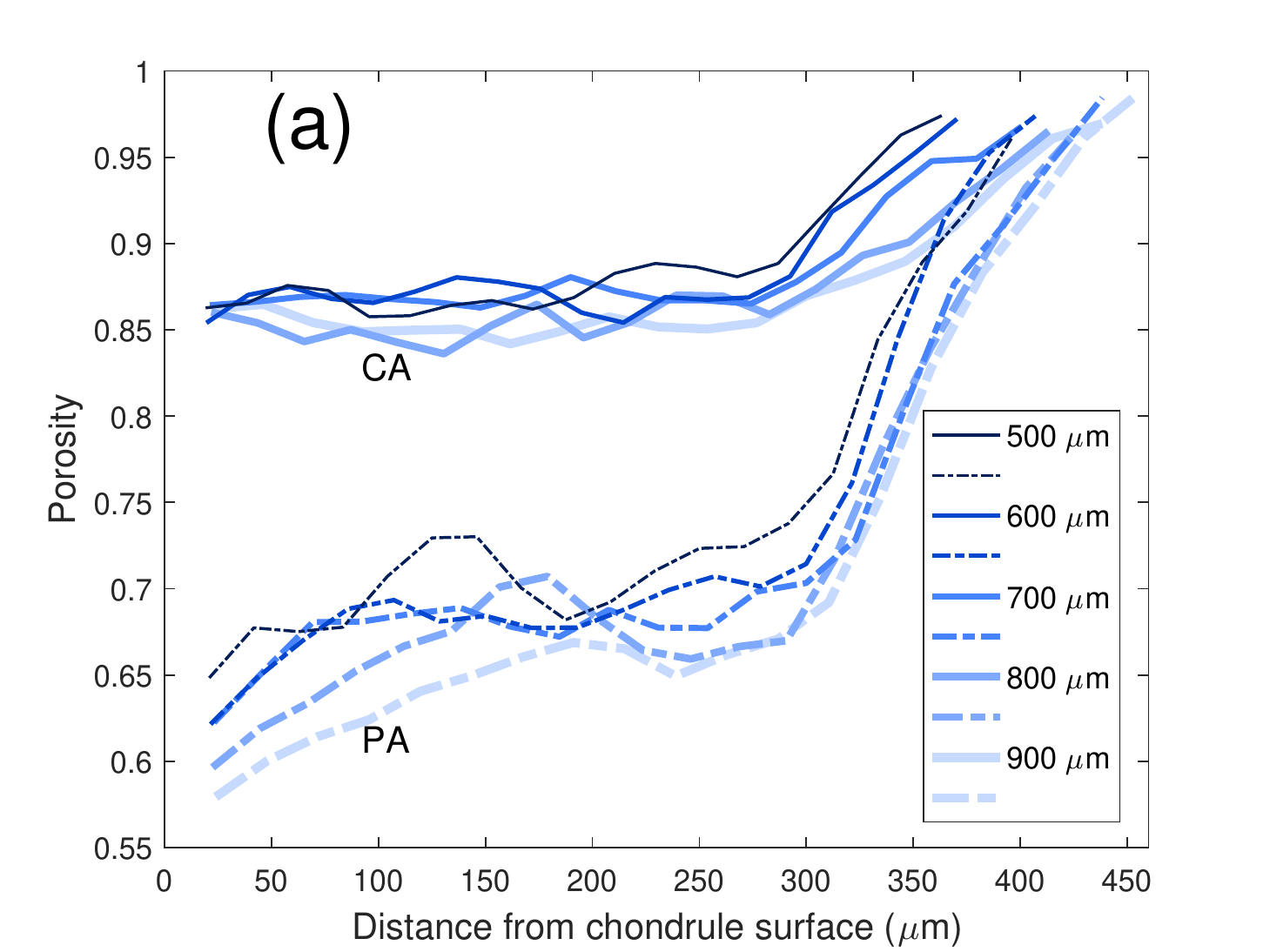}\includegraphics[width=9cm]{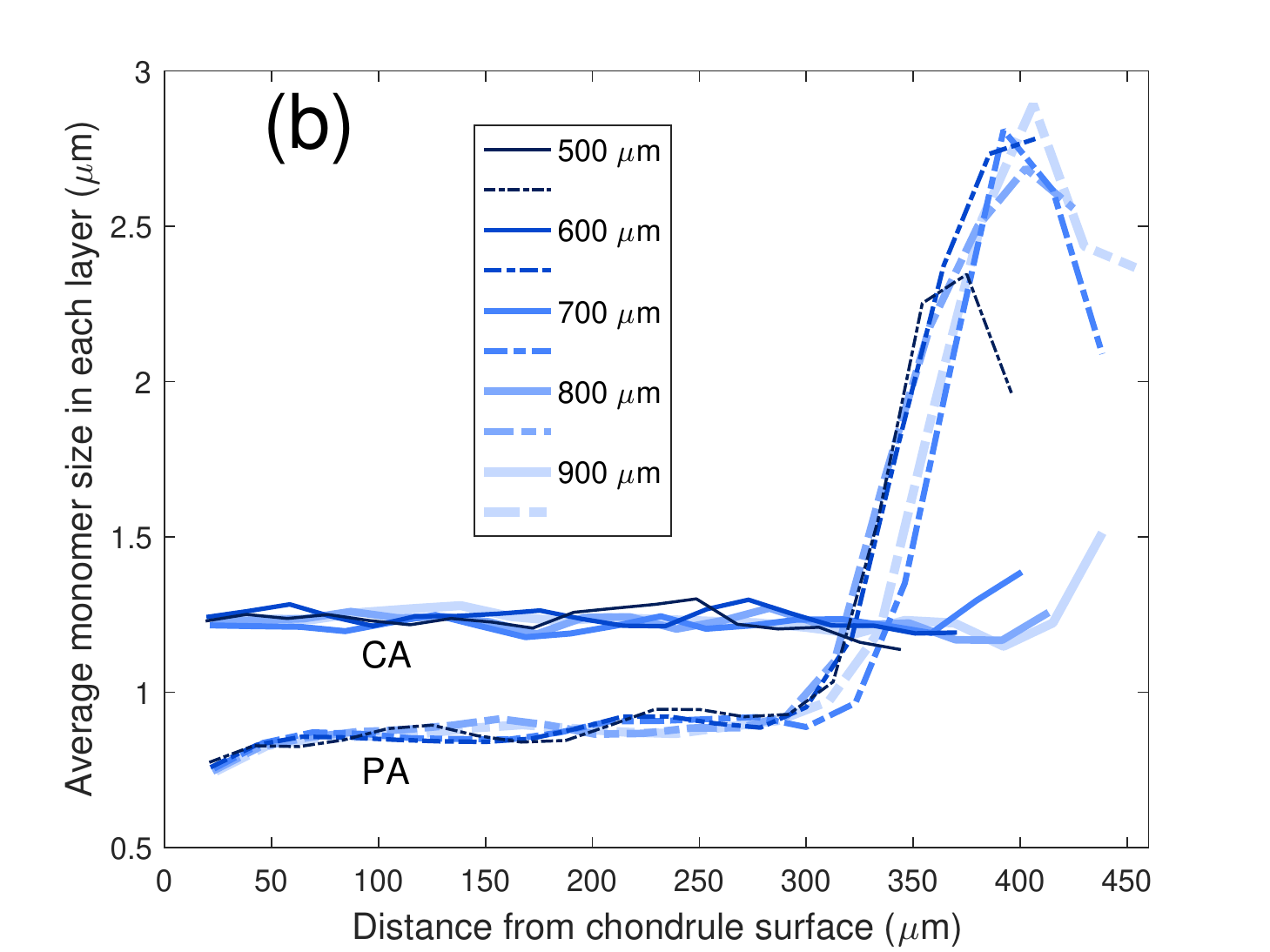}
\caption{Radial profiles of a) dust rim porosity and b) average monomer size in each horizontal layer on chondrules of different radii, for same rim thickness of  $\sim 290\mu m$ (the rim thickness is defined in Sec. \ref{sec:effturb}). Shown are data for particle aggregation (PA, runs a4-r$\ell$, $\ell=5, 6, ..., 9$) and cluster aggregation (CA, runs a4-r$\ell$-agg, $\ell=5, 6, ..., 9$).}
\label{fig:f13}
\end{figure*}

As dust particles in the solar nebula can form aggregates before colliding with chondrules (e.g., Scott et al. 1984), it is instructive to compare dust rims formed through accretion of single monomers (particle aggregation, PA) and accretion of aggregates (cluster aggregation, CA). A library of small aggregates was created by building aggregates from spherical grains with the same size distribution as the monomer library. The aggregates were built using a combination of PCA (particle-cluster aggregation) and CCA (cluster-cluster aggregation), and their equivalent radii range from $0.5 \mu m<R _{\sigma}< 10 \mu m$, while their physical radii range from $1 \mu m<R< 24 \mu m$, (see Figure. \ref{fig0} for the definition of $R _{\sigma}$ and $R$). The aggregates were binned by their equivalent radius, which was also used to calculate the relative velocity between the aggregate grain and chondrule (Eq. \ref{eq:vturb}). Since aggregates are more porous than monomers and have larger radii than monomers of the same mass, it is less likely that the aggregates in CA will pass though the gaps of the dust rim. Instead, they are likely to stick to the outer layers of the rim. Therefore, the dust rims formed by CA are more porous than for PA, as shown in Fig. \ref{f15} and \ref{fig:f13}a. Since almost no restructuring takes place in weak turbulence ($\alpha =10^{-4}$), the incoming particles simply add to the outer layer. Hence, the inner region of the rims for CA (up to $\sim 280\mu m$) has an approximate constancy of porosity as a function of distance from the chondrule surface, while for PA the porosity increases from the base of the rim to the top, with more small monomers filling the inner layers. 
In addition to the difference in porosity, Fig. \ref{fig:f13}b shows that the dust rims formed by CA have more constant monomer size distribution throughout the rims, while the average monomer size increases (especially at the top of the rims) with distance from chondrule center for dust rims formed by PA.

%%%%%%%%%%%%%%%%%%%%%%%%%%%%%%%%%%%%%%%%%%%%%%%%%%%%%%%%%%%%%
\section{Discussion}

We have calculated FGR porosities for different values of the parameters involved in the collision between chondrules and dust (either a spherical monomer or an aggregate; Secs. \ref{sec:effturb}-\ref{sec:effpaca}). We have also corroborated the linear relationship between FGR thickness and chondrule radius that has been measured by other authors (Sec. \ref{sec:effch}). We now put our results in the context of previous FGR studies.

\subsection{Structure and porosity of FGRs}

\begin{figure*}
\includegraphics[width=9cm,angle=-90]{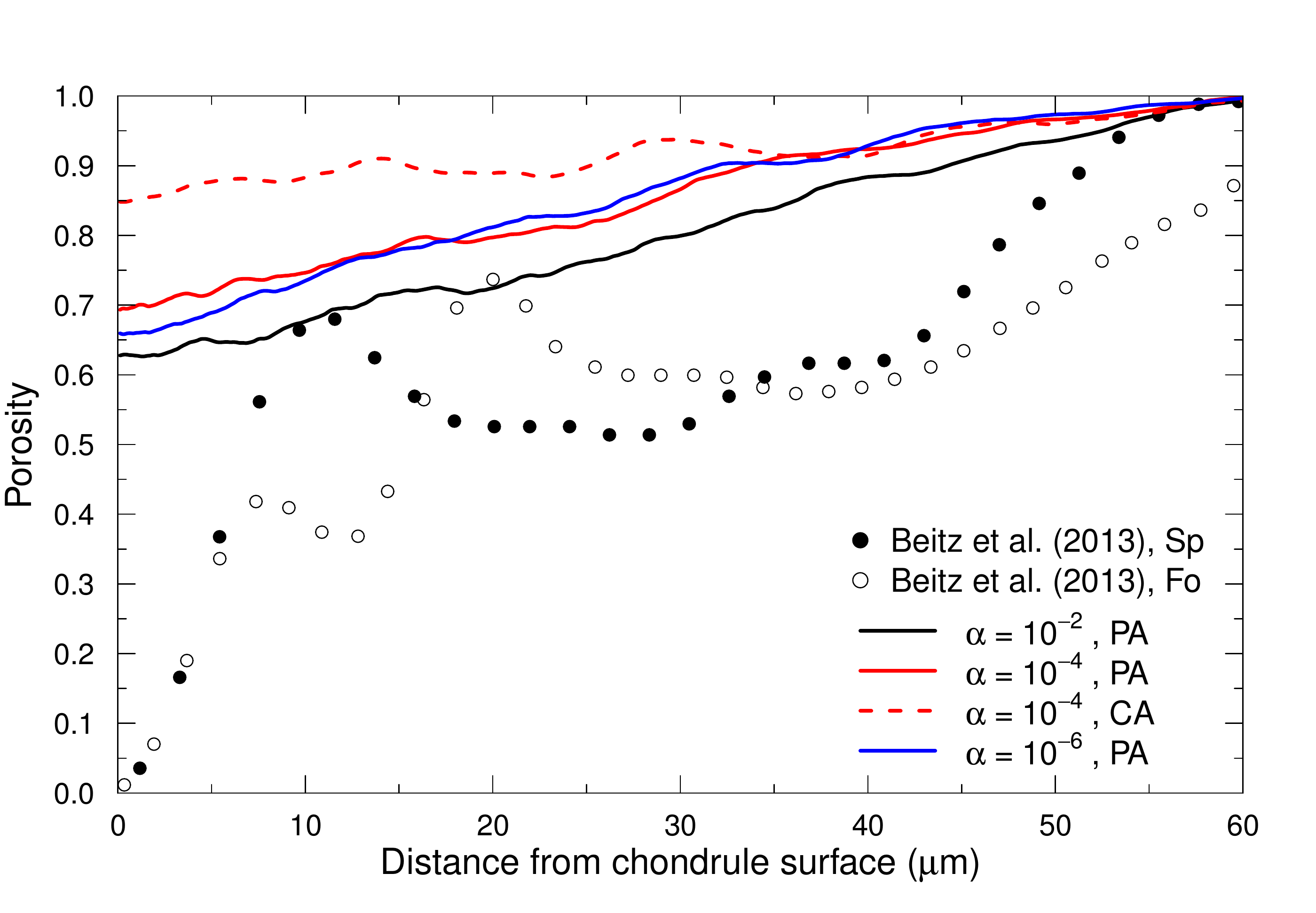}
\caption{Measurements of rim porosity obtained by Beitz et al. (2013) from experiments of rim accretion onto spinel (\textit{black points}) and forsterite (\textit{white points}) chondrule analogs. For comparison, data from our runs a2-r5 (\textit{black curve}), a4-r5 (\textit{red, solid curve}), a4-r5-agg (\textit{red, dashed curve}), and a6-r5 (\textit{blue curve}) are also shown. }
\label{fig:porosity_comparison}
\end{figure*}

Beitz et al. (2013) investigated FGR formation around chondrule analogs in laboratory experiments. The authors produced two types of chondrule analogs: one with a forsterite composition (radius $=0.75$ mm) and one with a spinel composition (radius $=0.80$ mm). The chondrule analogs were levitated inside a funnel using a gas flow, which also contained olivine dust grains of irregular shapes. These grains stuck to the chondrule analogs, with most of the stuck grains having radii in the range $\sim 0.25 - 1.5$ $\mu$m. Using scanning electron microscopy and X-ray computed tomography (CT), Beitz et al. (2013) measured the porosity profiles of the formed rims. Their data is shown in Fig. \ref{fig:porosity_comparison}: the \textit{black points} correspond to the rim porosity around the spinel chondrule analog, while the \textit{white points} represent rim porosity values around the forsterite chondrule analog. 

Figure \ref{fig:porosity_comparison} also shows our data for runs a2-r5 (\textit{black curve}), a4-r5 (\textit{red, solid curve}), a4-r5-agg (\textit{red, dashed curve}), and a6-r5 (\textit{blue curve}). Perhaps the largest difference between the numerical and the experimental data is seen close to the chondrule surface: the former shows a high porosity in the lower rim layers, while the latter exhibits very low porosity there. As Beitz et al. (2013) explain, the low porosity at the boundary between their chondrule analog and its associated dust rim is due to partial melting or sintering. Even if no melting or sintering occurred, we speculate that the low porosity close to a chondrule analog surface might be possible due to the irregular shape of the dust monomers, which can arrange themselves in more compact configurations than the spherical monomers used in our simulations. This difference in monomer shape is perhaps also responsible for the overall higher porosities in our simulations, compared to the experimentally obtained rims. Future simulations with non-spherical monomers will test these hypotheses. 

The high FGR porosities calculated from our simulations ($\gtrsim 60\%$), however, are generally consistent with those obtained by Ormel et al. (2008) for the dust component of chondrule aggregates ($\gtrsim 67\%$), as well as with the initial porosity estimated for an FGR in Allende (70-$80\%$, Bland et al. 2011). Such high porosities could conceivably result from gentle collisions between dust and chondrules in weak turbulence (Ormel et al. 2008).

% Note, however, that even for very strong turbulence ($\alpha=0.1$) we obtain rim porosities as high as $\sim 55\%$ (Fig. \ref{fig:f3}). 

Effects that will need to be taken into account in future work include fragmentation and erosion due to energetic collisions, as well as rim compaction due to chondrule-chondrule collisions.

\subsection{Comparison to FGR observations}

The ultimate goal of our modeling efforts is the comparison of our results to observations of FGRs in chondrites to help shed light on the nebular conditions of FGR formation.  However, the calculations presented herein apply only to the \textit{initial} accretion of FGRs, and do not provide information regarding their subsequent structural evolution. For example, the role of FGRs as the ``glue" that holds together decimeter-sized chondrule composites (Ormel et al. 2008) means that rims would be subjected to compaction events via collisions between rimmed chondrules and other, perhaps similar, objects. It has also been suggested that weak nebular shocks may restructure the rim grains leading to a compaction fabric and lower rim porosity (Bland et al. 2011).  After accretion to the chondrite parent body, the rims may be modified further by impact compaction, thermal metamorphism, and aqueous alteration.  All of these processes will likely modify the rim structure and reduce porosity.  Regardless, modeling the initial accretion of FGRs is an important first step towards identifying which of these nebular and parent body processes may have been involved in, and their relative influence on, post-formation FGR modification.

Bland et al. (2011) examined a CV Allende FGR fabric defined by the crystallographic alignment of olivine grains within the rim.  By estimating the compressional strain needed to produce such a fabric and combining this with the current Allende porosity measured by Consolmagno et al. (2008), they estimated an initial rim porosity on the order of 70-80\%. They interpreted the compaction as a post-accretional nebular process, such as a nebular shock or rimmed chondrule collision, so this initial porosity estimate should be comparable to our modeled rim porosity.  For our model, porosities this high are associated with smaller chondrules (in the particle aggregation case) or with cluster aggregation (Fig. \ref{fig:f13}).  Because the Allende chondrule is over $\sim 1$ mm in size, this seems to exclude particle aggregation for the formation of the rim unless the turbulence strength was much lower than $10^{-4}$ (Fig. \ref{fig:f3}).  One important difference between our model and the Allende FGR, however, is that our model uses perfectly spherical grains while olivine grains are typically elongated.  As we hypothesize above, irregular monomer shape may decrease the rim porosity, but future work will investigate this.

Another study that characterized strain in FGRs among several chondrules in CM Murchison found that the compression of the rims likely occurred on the parent body (Hanna and Ketcham, 2018).  In this case, the strain estimate from compression of the rims and the current bulk porosity of Murchison leads to a pre-deformation rim porosity of 45\% (Appendix C). Because the porosity reduction took place on the parent body it is difficult to know if this porosity estimate is directly comparable to our modeling results.  In light of the Bland et al. (2011) study showing evidence of pre-parent body compression of the rim, it is likely that the 45\% estimate does not represent original FGR formation porosity.  Another complication is that FGR rims in CM chondrites are dominated by serpentine (a hydrous mineral) in contrast to the anhydrous olivine dominant in CV FGRs (Zolensky et al., 1993). Therefore, if CM FGRs were aqueously altered after formation, this also would have reduced the original formation porosity [although whether this hydration took place in the nebula or on the parent body is debated (e.g., Haenecour et al., 2018; Metzler et al., 1992; Tomeoka and Tanimura, 2000)].  Still, these data place a lower limit ($\sim 45\%$) on the original, formational FGR porosity and is consistent with our modeling results that suggest a minimum original porosity of $\sim 55\%$.

There have been two direct observations of current, in-situ porosities of FGRs in carbonaceous chondrites.  Beitz et al. (2013) calculated a ~10\% porosity for two CM Murchison rims from X-ray CT data. However, their assumptions of zero porosity in chondrule interiors and an identical mineralogical and chemical composition of rims and chondrules are likely incorrect (e.g., Fuchs et al., 1973; Hanna et al., 2015). Additionally, repeated measurements of Murchison porosity using helium ideal-gas pycnometry indicates a much higher porosity of 22.1\% (Macke et al., 2011).  Haenecour et al. (2018) also estimated a ~10\% porosity for FGRs in two primitive CO3.0 chondrites from reduced analytical totals in electron microprobe analysis (EMPA). This FGR porosity matched both the average matrix value and the average porosity of CO chondrites [10.8\%; (Consolmagno et al., 2008)].  This observed porosity is much lower than our modeled porosity and may again represent post-accretional processing of the FGR, most likely aqueous alteration which the authors found evidence of within the FGRs.

Finally, we note that another assumption in our model is that the chondrule surface is smooth and spherical.  In reality, chondrule surfaces can be rough and chondrule shapes can depart from spherical, possibly as primary features (Hanna \& Ketcham 2018).  Comparison of FGR volume with chondrule roughness suggests that increased chondrule surface roughness caused greater accumulation of dust onto the chondrule surface (Hanna and Ketcham, 2018).  Forthcoming work will examine the influence of irregular chondrule surface shape on dust accumulation in our model.

\subsection{FGR formation times}
The results of Fig. \ref{fig:f3_2} for the formation times of FGRs of a certain thickness (40, 180 and 320 $\mu$m) can be qualitatively compared to previous estimates by other authors. Cuzzi (2004) showed that rimming times, like our data in Fig. \ref{fig:f3_2}, decrease with increasing $\alpha$. However, the rimming times calculated by Cuzzi (2004) correspond to a solar nebula location of 2.5 AU, where turbulent velocities are lower than at 1 AU as we consider here.  Depending on the ratio $\zeta$ of rim volume to chondrule volume, rimming times obtained by Cuzzi (2004) vary between 40 yr ($\zeta=0.1$) and 600 yr ($\zeta=3.0$) at 2.5 AU, whereas at 1 AU we find the rimming rimes to be between 0.3 yr ($\zeta=0.1$) and 10 yr ($\zeta=3.0$), for $\alpha=10^{-4}$. 

Ormel et al. (2008) calculated the times at which the available dust in their simulations was depleted by incorporation onto chondrule surfaces, as a function of $\alpha$. Once again, those times decrease with increasing $\alpha$. As Ormel et al. (2008) take into account the growth of chondrule aggregates (i.e., objects composed of chondrules joined by fine-grained dust), direct comparison to our rimming times is difficult.  

Gunkelmann et al. (2017) point out that the bouncing velocity of two chondrules increases by two orders of magnitude if they are covered by dust rims. Thicker and denser dust rims are more efficient in accommodating the collision energy (with a higher bouncing velocity) than thin and porous rims, while the dust rims are partly destroyed in all cases by sputtering, even for hit-and-stick events. Our simulations show that the dust rims formed in different turbulence conditions have different porosity and growth rates, which means the colliding chondrules in different environments will have different bouncing velocities, affecting the growth of chondrule agglomerates. It will be of interest to investigate the growth rates of chondrules in various turbulent environments by conducting collisions between rimmed chondrules in our future work.

%As chondrules sweep up the local dust particles and grow in size, they experience higher relative velocities, which can ultimately exceed the threshold fragmentation velocity (above $\sim m s^{-1}$) (Ormel et al., 2008). It's unlikely that the collisions between chondrules and the dust with high relative velocities can cause fragmentation because of the low mass ratio between them, but rather a more possible outcome is the local erosion (Krijt, S.2015). However, for collisions between chondrules with high kinetic energy, they are very likely to compress the existing dust rims or fragment. The compaction of the dust rims results in a longer friction time of the chondrule and therefore a even higher relative velocity, which can cause more frequent bouncing, erosion or fragmentation, limiting the the further growth of the dust rims. The compaction or disruption of the dust rims also affect their charge, due to the change of the surface areas. The variation of charge can alter the ionization state of the local environment and further impact the turbulence strength, which in turn affects the collisions and dust rim growth. For a more complete model, it is necessary to take into account the erosion, fragmentation, as well as the compaction of the dust rims through collisions between chondrules in our future work. Also it may be instructive to analyze the relationships between the environmental parameters and rim structures in the steady state where the dust rims stops growing.

%%%%%%%%%%%%%%%%%%%%%%%%%%%%%%%%%%%%%%%%%%%%%%%%%%%%%%%%%%%%%
\section{Conclusions}

We conducted a numerical study of the initial accretion of fine-grained dust rims (FGRs) onto chondrule surfaces. This study is an important first step towards elucidating the structural properties of FGRs \textit{before} the onset of impact compaction, thermal metamorphism, and aqueous alteration. 

In this work, we concentrated on the porosity and the thickness of FGRs as signatures of the collisional formation process of rim structures. We compared FGRs formed in nebular turbulence of different strengths, and also investigated the roles of chondrule size and cluster aggregation. Our main conclusions are:

%We have presented a model that incorporates the detailed physical characteristics of aggregates in the collision process and used MC algorithm to model the time evolution
%of the dust rims in a protoplanetary disk. We defined two quantities, the porosity
%and the thickness, to quantify the effect of the collision process on the structure of the dust rim. We compared the dust rims formed in different turbulence strengths, and also investigated the impact of the chondrule size as well as the porosity of the dust particles on the rim formations. Our main conclusions are:

%1. Most collisions in low turbulence result in little restructuring, and therefore hit-and-stick is the main process of dust rim growth for small particles and low turbulence strength ($\alpha \leq 10^{-4}$).

1. FGRs formed in environments with strong turbulence are more compact and grow more rapidly than FGRs formed in weak turbulence. In the case of strong turbulence ($\alpha \gtrsim 10^{-2}$), FGR porosity has an approximately constant value of 55 -- 60\% in the inner regions of the rims, while in the case of weak turbulence ($\alpha \lesssim 10^{-3}$) the porosity increases with distance from the chondrule center, from values of $\sim 60\%$ to $\sim 70\%$.

2. The times needed to build FGRs of a certain thickness decrease approximately linearly with increasing turbulent parameter $\alpha$. 

3. FGR thickness scales linearly with chondrule radius. This linear relation becomes steeper over time, as the greater collision cross section and higher relative velocities of the larger chondrules increase the collision rate.

4. The mean porosity of FGRs decreases over time in weak turbulence ($\alpha=10^{-4}$) for all chondrule sizes studied. Compaction reduces porosity from early values of $\sim 85\%$ to values between $\sim 66$ -- $68\%$ in $\sim 100$ years. 

5. FGRs formed by cluster aggregation have higher porosity ($\sim 85\%$) than those formed by accretion of individual monomers ($\sim 60$--$70\%$), for all chondrule sizes studied. 
\\
%\\[8pt]

\textbf{Acknowledgments:} this work was supported by the National Science Foundation under grant no. 1414523 (LSM and TWH).

%Other factors, such as thermal/aqueous and nebula shock waves, that may lead to a lower porosity are not considered in this model. Future work should take these factors into consideration, and investigate the effect of the dust rim on the collisions between chondrules.\\

\appendix

\section*{Appendices}
\addcontentsline{toc}{section}{Appendices}
\renewcommand{\thesubsection}{\Alph{subsection}}
\subsection{Energy dissipation mechanism during restructuring}

\begin{figure*}
\includegraphics[width=9cm]{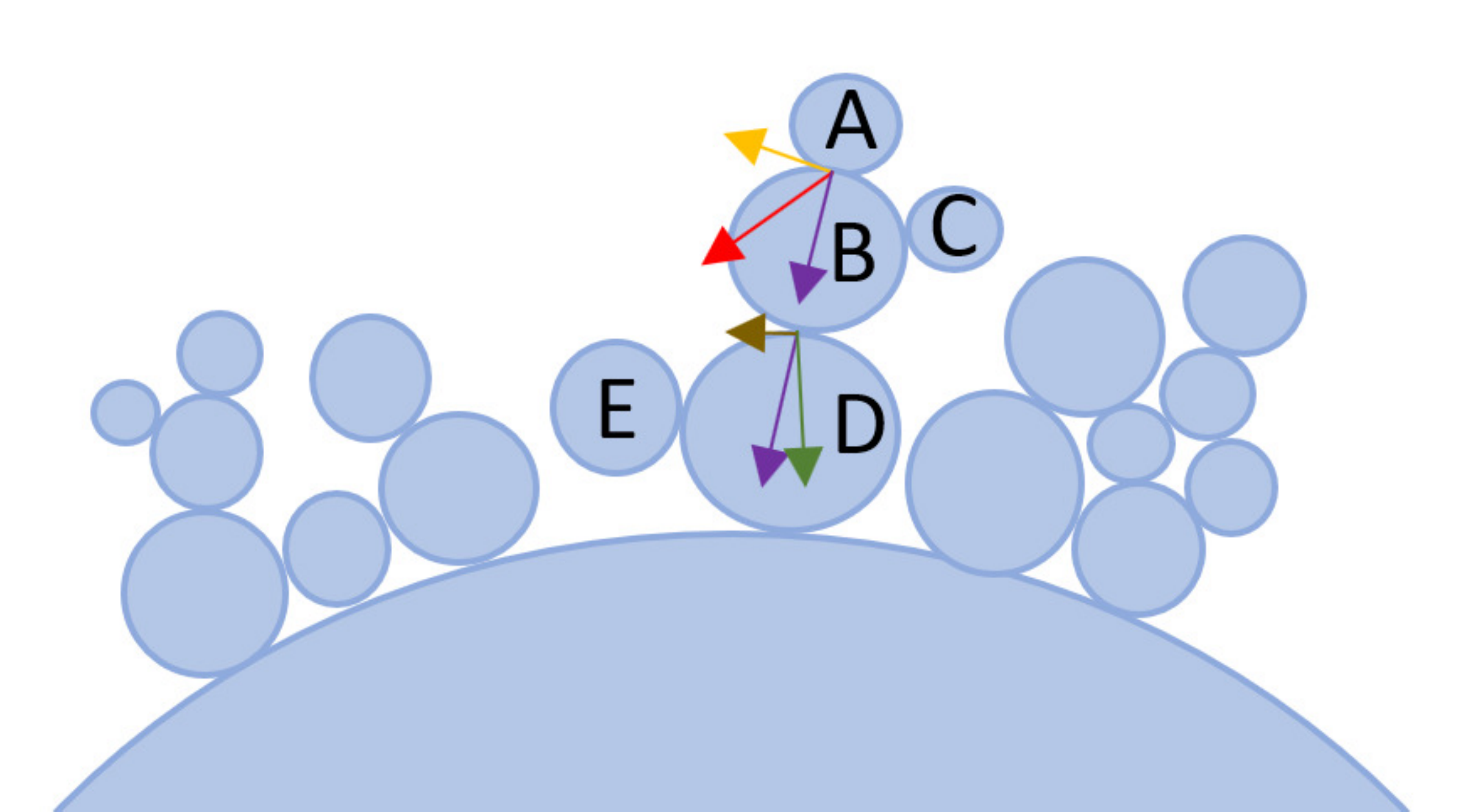}
\caption{Illustration of momentum transfer within the dust rim. When the incoming grain A hits grain B, its momentum (red) is decomposed into the tangential component (yellow) which makes grain A roll counterclockwise on the surface of grain B and the normal component (purple) which is transferred to grain B. The momentum of grain B gets decomposed again at the contact point between grain B and grain D. The tangential component (brown) enables grains B, A and C roll counterclockwise as a whole on the surface of D, and the normal component is transferred to grain D.}
\label{fig15}
\end{figure*}

The momentum of the incoming particle is decomposed into directions tangential and perpendicular to the point of contact. The kinetic energy associated with the tangential component of the momentum is used for the rolling of the incoming particle (if it exceeds the critical rolling energy). The perpendicular component is transferred to the sphere in contact with the incoming particle, and is decomposed again at its contact point with another sphere. In this manner the momentum of the incoming particle is transferred to the inner spheres of the dust rim, and the energy of $e_{roll}$ is dissipated as a sphere rolls a distance of 1 $\AA$. The momentum stops transferring inward when the remaining energy is not sufficient to cause any rolling ($E< e_{roll}$). Figure \ref{fig15} illustrates the process of momentum transfer. Note that the grains don’t necessarily stop rolling simultaneously, as each sub system has different kinetic energy. For example, grain A may stop rolling on grain B due to the depletion of its tangential momentum but may still roll together with grain B and C on top of grain D.

%%%%%%%%%%%%%%%%%%%%%%%%%%%%%%%%%%%%%%%%%%%%%%%%%%%%%%%%%%%%%

\subsection{Physical meaning of non-integer weight}
The physical meaning of a non-integer weight can be interpreted by expanding the computational volume in which the dust particles reside. Suppose the weight of the colliding particle is 0.2; the collision can be imagined to take place in a space which is five times as large as the original one, so that there is one such particle in the enlarged volume, and meanwhile, the weight of each species, including the chondrule, becomes five times as large. Therefore, the probability that particle $d$ will collide with the chondrule is: $C_{ch,d}(new)=5 \times 5w_{j}(old)\sigma _{ch,d}\Delta v_{ch,d}/5V=5C_{ch,d}(old)$, where ‘old’ refers to the original space and ‘new’ refers to the enlarged space. Since the chance that the collision takes place in the original space is 1/5, the probability that particle $d$ will collide with the chondrule in the original space is $C_{ch,d}(new)/5$, which equals to $C_{ch,d}(old)$. 

%%%%%%%%%%%%%%%%%%%%%%%%%%%%%%%%%%%%%%%%%%%%%%%%%%%%%%%%%%%%%
\subsection{Initial porosity of compacted Murchison FGRs}
Hanna and Ketcham  (2018) found that that Murchison FGRs varied in thickness and were consistently thinner (compressed) in the direction of maximum strain within the meteorite.  The maximum FGR thickness was 9.2\% of the maximum chondrule length and the minimum FGR thickness was 6.4\% (of maximum chondrule length) [refer to Fig. 6F of Hanna and Ketcham (2018)].  Therefore the maximum relative thickness difference between compressed and uncompressed portions of the rim is 0.064/0.092 = 0.70.  In other words, the thickness of the FGRs has been reduced by up to 30\%.  If we assume the current bulk porosity of Murchison (22.1\%; Macke et al., 2011) is the current porosity of the FGRs in Murchison, we can use equation (9) of Hanna et al. (2015) to estimate the maximum initial porosity of the FGRs at $\sim$45\%.

%%%%%%%%%%%%%%%%%%%%%%%%%%%%%%%%%%%%%%%%%%%%%%%%%%%%%%

%\begin{thebibliography}{9}

\def\bibindent{1em}

\end{document}